\documentclass[journal]{IEEEtran}
\usepackage{blindtext}
\usepackage[utf8]{inputenc}
\usepackage[T1]{fontenc}
\usepackage[noadjust]{cite}
\usepackage{graphicx}
\usepackage[font=footnotesize]{caption}
\usepackage{dblfloatfix}
\usepackage{psfrag}
\usepackage{subcaption}
\usepackage{amsmath,amssymb,amsthm,mathrsfs,amsfonts,dsfont}
\usepackage{color}
\usepackage[algoruled]{algorithm2e}
\usepackage{fixltx2e}
\usepackage{multirow}
\usepackage{multicol}
\usepackage[normalem]{ulem}
\usepackage{acronym}
\usepackage[table,xcdraw]{xcolor}
\usepackage{accents}
\usepackage{mwe}
\usepackage{hhline}
\usepackage{trfsigns}
\usepackage{siunitx}
\usepackage{slashbox}
\usepackage{wasysym}
\usepackage{tikz}
\usepackage{booktabs}
\usetikzlibrary{shapes.geometric}
\usetikzlibrary{math}
\usetikzlibrary{backgrounds}
\usetikzlibrary{patterns}
\usetikzlibrary{positioning}
\usetikzlibrary{snakes}
\usetikzlibrary{calc,angles}
\definecolor{1}{rgb}{0.049031372549020,0.456884313725490,0.582331372549020}
\definecolor{2}{rgb}{0.847058823529412,0.450980392156863,0.498039215686275}
\definecolor{3}{rgb}{0.988235294117647,0.733333333333333,0.427450980392157}

\newacro{5G}{fifth generation}
\newacro{6G}{sixth generation}
\newacro{A/D}{analog-to-digital}
\newacro{ACDC}{all-cell Doppler correction}
\newacro{ADC}{analog-to-digital converter}
\newacro{AWGN}{additive white Gaussian noise}
\newacro{BB}{baseband}
\newacro{BER}{bit error ratio}
\newacro{BPSK}{binary phase-shift keying}
\newacro{BP}{band-pass}
\newacro{CCDF}{complementary cumulative distribution function}
\newacro{CDM}{code-division multiplexing}
\newacro{CFO}{carrier frequency offset}
\newacro{CFR}{channel frequency response}
\newacro{CIR}{channel impulse response}
\newacro{CP}{cyclic prefix}
\newacro{CPO}{carrier phase offset}
\newacro{CP-DSSS}{cyclic-prefix direct-sequence spread spectrum}
\newacro{CR}[C\&R]{wireless communication and radar sensing}
\newacro{CS}{chirp sequence}
\newacro{CW}{continuous wave}
\newacro{C-OCDM}{conventional OCDM}
\newacro{D/A}{digital-to-analog}
\newacro{DAC}{digital-to-analog converter}
\newacro{DDS}{direct digital synthesis}
\newacro{DFCS}{dual-functional communication and radar sensing}
\newacro{DFRC}{dual-functional radar-communication}
\newacro{DFnT}{discrete Fresnel transform}
\newacro{DFT}{discrete Fourier transform}
\newacro{DMRS}{demodulation reference signal}
\newacro{DoA}{direction of arrival}
\newacro{EVM}{error vector magnitude}
\newacro{FBMC}{filter bank multicarrier}
\newacro{FDE}{frequency-domain equalization}
\newacro{FDM}{frequency-division multiplexing}
\newacro{FDZP}{frequency-domain zero padding}
\newacro{FMCW}{frequency-modulated continuous wave}
\newacro{FrDM}{Fresnel-division multiplexing}
\newacro{FSP}{frequency-shift precoding}
\newacro{GFDM}{generalized frequency-division multiplexing}
\newacro{HAD}{highly automated driving}
\newacro{HP}{high-pass}
\newacro{IBFD}{in-band full duplex}
\newacro{IC}{interference cancellation}
\newacro{ICI}{intercarrier interference}
\newacro{IChI}{interchirp interference}
\newacro{IDFT}{inverse discrete Fourier transform}
\newacro{IDFnT}{inverse discrete Fresnel transform}
\newacro{IF}{intermediate frequency}
\newacro{ISAC}{integrated sensing and communication}
\newacro{ISI}{intersymbol interference}
\newacro{ISLR}{integrated-sidelobe level ratio}
\newacro{JCAS}{joint communication and sensing}
\newacro{JRC}{joint radar-communications}
\newacro{LNA}{low-noise amplifier}
\newacro{LO}{local oscillator}
\newacro{LoS}{line-of-sight}
\newacro{LP}{low-pass}
\newacro{LFSR}{linear-feedback shift register}
\newacro{mmWave}{milimeter wave}
\newacro{MIMO}{multiple-input multiple-output}
\newacro{MLS}{maximum-length sequence}
\newacro{MMSE}{minimum mean square error}
\newacro{MU}{multi-user}
\newacro{NBI}{narrowband interference}
\newacro{NLoS}{non-line-of-sight}
\newacro{NR}{New Radio}
\newacro{NRMSE}{normalized root mean square error}
\newacro{OCDM}{orthogonal chirp-division multiplexing}
\newacro{OFDM}{orthogonal frequency-division multiplexing}
\newacro{OOB}{out-of-band}
\newacro{OTFS}{orthogonal time-frequency space}
\newacro{P/S}{parallel-to-serial}
\newacro{PA}{power amplifier}
\newacro{PACF}{periodic autocorrelation function}
\newacro{PAPR}{peak-to-average power ratio}
\newacro{PCCF}{periodic cross-correlation function}
\newacro{PLC}{powerline communication}
\newacro{PLL}{phase-locked loop}
\newacro{PMCW}{phase-modulated continuous wave}
\newacro{PMEPR}{peak-to-mean envelope power ratio}
\newacro{PPLR}{peak power loss ratio}
\newacro{PRBS}{pseudorandom binary sequence}
\newacro{PSLR}{peak sidelobe level ratio}
\newacro{QPSK}{quadrature phase-shift keying}
\newacro{RaaS}{radar as a service}
\newacro{RadCom}{radar-communication}
\newacro{RCS}{radar cross section}
\newacro{RF}{radio-frequency}
\newacro{RTS}{radar target simulator}
\newacro{SDM}{spatial division multiplexing}
\newacro{SDMA}{spatial division multiple access}
\newacro{SH}[S\&H]{sample and hold}
\newacro{SI}{self-interference}
\newacro{SIC}{self-interference cancellation}
\newacro{SISO}{single-input single-output}
\newacro{SM-OCDM}{sector-modulated OCDM}
\newacro{SNR}{signal-to-noise ratio}
\newacro{S/P}{serial-to-parallel}
\newacro{SC}[S\&C]{Schmidl \& Cox}
\newacro{SoC}{system-on-a-chip}
\newacro{TDD}{time-division duplexing}
\newacro{TDE}{time-domain equalization}
\newacro{TDM}{time-division multiplexing}
\newacro{TDR}{time-domain reflectometry}
\newacro{ToF}{time of flight}
\newacro{UAV}{unmanned aerial vehicle}
\newacro{UE}{user equipment}
\newacro{UWAC}{underwater acoustic communication}
\newacro{V2I}{vehicle-to-infrastructure}
\newacro{V2V}{vehicle-to-vehicle}
\newacro{ZCZ}{zero correlation zone}
\newacro{ZF}{zero forcing}
\newacro{ZP}{zero padding}

\newcommand{\SPup}{0.7,1.5}
\newcommand{\SPdown}{-0.7,-1.5}

\tikzstyle{box1} = [rectangle,draw,text = black,fill = gray!20,line width=0.5mm,rotate=0, node distance=6.28em, text centered, minimum height=2em,  minimum width=3.5em]
\tikzstyle{box2} = [rectangle,draw,text = black,fill = gray!20,line width=0.5mm,rotate=0, node distance=5.27em, text centered, minimum height=2em,  minimum width=3.5em]
\tikzstyle{box3} = [rectangle,draw,text = black,fill = gray!20,line width=0.5mm,rotate=0, node distance=4.52em, text centered, minimum height=4em,  minimum width=2em]
\tikzstyle{box4} = [rectangle,draw,text = black,fill = gray!20,line width=0.5mm,rotate=0, node distance=4.52em, text centered, minimum height=2em,  minimum width=3.5em]
\tikzstyle{box5} = [rectangle,draw,text = black,fill = gray!20,line width=0.5mm,rotate=0, node distance=4em, text centered, minimum height=2em,  minimum width=3.2em]
\tikzstyle{box6} = [rectangle,draw,text = black,fill = gray!20,line width=0.5mm,rotate=0, node distance=3em, text centered, minimum height=2em,  minimum width=2em]

\tikzstyle{boxSynch} = [rectangle,draw,text = black,fill = gray!20,line width=0.5mm,rotate=0, node distance=7.3em, text centered, minimum height=2em,  minimum width=4.7em]
\tikzstyle{boxSuC} = [rectangle,draw,text = black,fill = gray!20,line width=0.5mm,rotate=0, node distance=5.02em, text centered, minimum height=2em,  minimum width=6.5em]

\makeatletter
\renewcommand*\env@cases[1][1.2]{%
	\let\@ifnextchar\new@ifnextchar
	\left\lbrace
	\def\arraystretch{#1}%
	\array{@{}l@{\quad}l@{}}%
}
\makeatother

\usepackage{fancyhdr}
\fancypagestyle{empty}{fancy}

\begin{document}
	
\title{Bistatic OFDM-based ISAC with Over-the-Air Synchronization: System Concept and\\ Performance Analysis}

\author{David Brunner,
	Lucas Giroto de Oliveira,~\IEEEmembership{Graduate Student Member,~IEEE},\\ Charlotte Muth,~\IEEEmembership{Graduate Student Member,~IEEE}, Silvio Mandelli,~\IEEEmembership{Member,~IEEE},\\ Marcus Henninger,~\IEEEmembership{Member,~IEEE}, Axel Diewald, Yueheng Li,\\ Mohamad Basim Alabd,~\IEEEmembership{Graduate Student Member,~IEEE}, Laurent Schmalen,~\IEEEmembership{Fellow,~IEEE},\\ Thomas Zwick,~\IEEEmembership{Fellow,~IEEE}, and Benjamin Nuss,~\IEEEmembership{Senior Member,~IEEE}
	\thanks{\iffalse Manuscript received MONTH DAY, YEAR; revised November, 14, 2022; accepted November 19, 2022\fi. The authors acknowledge the financial support by the Federal Ministry of Education and Research of Germany in the projects ``KOMSENS-6G'' (grant number: 16KISK123) and ``Open6GHub'' (grant number: 16KISK010). \textit{(Corresponding author: Lucas Giroto de Oliveira.)}}
	\thanks{D. Brunner was with the Institute of Radio Frequency Engineering and Electronics (IHE), Karlsruhe Institute of Technology (KIT), 76131 Karlsruhe, Germany. He is now with VEGA Grieshaber KG, 76137 Karlsruhe, Germany (e-mail: {d.brunner@vega.com}).}
	\thanks{L. Giroto de Oliveira, Y. Li, T. Zwick, and B. Nuss are with the Institute of Radio Frequency Engineering and Electronics (IHE), Karlsruhe Institute of Technology (KIT), 76131 Karlsruhe, Germany (e-mail: {lucas.oliveira@kit.edu}, {yueheng.li@kit.edu}, {thomas.zwick@kit.edu}, and {benjamin.nuss@kit.edu}).}
	\thanks{C. Muth and L. Schmalen are with the Communications Engineering Laboratory (CEL), Karlsruhe Institute of Technology (KIT), 76131 Karlsruhe, Germany (e-mail: charlotte.muth@kit.edu and laurent.schmalen@kit.edu)}
	\thanks{S. Mandelli and M. Henninger are with Nokia Bell Laboratories, 70469 Stuttgart, Germany (e-mail: silvio.mandelli@nokia-bell-labs.com and marcus.henninger@nokia.com).}
	\thanks{A. Diewald was with the with the Institute of Radio Frequency Engineering and Electronics (IHE), Karlsruhe Institute of Technology (KIT), 76131 Karlsruhe, Germany. He is now with Bruker BioSpin, 76275 Ettlingen, Germany (e-mail: diewald-axel@web.de).}
	\thanks{M. B. Alabd was with the with the Institute of Radio Frequency Engineering and Electronics (IHE), Karlsruhe Institute of Technology (KIT), 76131 Karlsruhe, Germany. He is now with IPG Automotive GmbH, 76185 Karlsruhe, Germany (e-mail: basim.alabd@ipg-automotive.com).}	
}


\maketitle

\begin{abstract}
	Integrated sensing and communication (ISAC) has been defined as one goal for 6G mobile communication systems. In this context, this article introduces a bistatic ISAC system based on orthogonal frequency-division multiplexing (OFDM). While the bistatic architecture brings advantages such as not demanding full duplex operation with respect to the monostatic one, the need for synchronizing transmitter and receiver is imposed. In this context, this article introuces a bistatic ISAC signal processing framework where an incoming OFDM-based ISAC signal undergoes over-the-air synchronization based on preamble symbols and pilots. Afterwards, bistatic radar processing is performed using either only pilot subcarriers or the full OFDM frame. The latter approach requires estimation of the originally transmitted frame based on communication processing and therefore error-free communication, which can be achieved via appropriate channel coding. The performance and limitations of the introduced system based on both aforementioned approaches are assessed via an analysis of the impact of residual synchronization mismatches and data decoding failures on both communication and radar performances. Finally, the performed analyses are validated by proof-of-concept measurement results.
\end{abstract}


\IEEEpeerreviewmaketitle

\section{Introduction}\label{sec:Introduction}

\IEEEPARstart{O}{ne} vision for 6G mobile communication systems is creating a digital representation of the physical world. Therefore, future communication systems must be capable of transmitting high data rates as well as precisely sensing their environment \cite{BBFKOMSENS6G}. For this purpose, appropriate sensor technology, such as radar technology, must be integrated into the communication system, constituting a integrated sensing and communication (ISAC) system \cite{Liu2023}. Future applications of ISAC range from indoor factory safety, roadway monitoring, to flying drone monitoring and detection \cite{Kadelka,NokiaPaperneu}. A further possible application of the ISAC concept are modern autonomous driving systems, which require rapid data exchange in vehicle-to-vehicle (V2V) and vehicle-to-infrastructure (V2I) communication, e.g., for collision avoidance, traffic coordination, and interference avoidance among automotive radar sensors \cite{Giroto,PaperAutonomesFahren}.

The waveform of the transmit signal is based on orthogonal frequency-division multiplexing (OFDM), as it is currently used in cellular 5G systems and will likely also be used in 6G systems \cite{NokiaPaperneu,Ksiezyk}. OFDM has proven to be very robust against inter symbol interference (ISI) in multipath propagation and provides excellent data transmission characteristics \cite{NokiaPaperneu}. Moreover, it offers further advantages through easy synchronization and equalization \cite{PaperSturm}. Since OFDM is also suitable as a digital modulation scheme for radar applications, it has been widely investigated for ISAC systems \cite{Giroto}.

While monostatic sensing is the mainly investigated deployment in the literature, the focus in this article is shifted to bistatic transmit and receive units, as full duplex hardware \cite{barneto2019,barneto2021}, which is not envisioned for communication and imposes higher costs as well as the need to deal with self-interference, is not required for bistatic sensing \cite{Thomae}. In this context, the functionality of a bistatic radar based on OFDM is demonstrated in \cite{Falcone} and \cite{Yildirim}. While these articles demonstrate the feasibility of bistatic sensing based on passive radar processing of OFDM-based WiFi signals, which requires synchronization is  between  transmitter (Tx) and receiver (Rx) to be performed, no detailed discussion on synchronization mismatches was made. In \cite{Smaini}, initial effects of synchronization offsets on the communication performance using the WiFi (802.11a/g) and WiMAX (802.16e) standards as examples. Fundamental investigations on the effects of synchronization offsets on radar can be found in \cite{PaperJUMP}. However, there is a lack of detailed considerations of practically relevant performance parameters of the radar. To the best of the authors' knowledge, there exists no publication or measurement-based demonstration of a bistatic OFDM-based ISAC system considering relevant practical aspects such as the combination of the OFDM frame structure, synchronization methodology, and the reconstruction of the transmitted frame based on channel-coded communication. An initial study of the presented ISAC system has been performed in \cite{unserPaper}. Extending the authors' previous work, this article presents a more detailed formulation of the ISAC system concept and a more systematic analysis of residual synchronization offsets, as well as over-the-air (OTA) measurements at \SI{79}{\giga\hertz}.

In this context, this article deals with the design and the investigation of a bistatic OFDM-based ISAC system. The main contributions are as follows:
\begin{itemize}

	\item A mathematical formulation of the considered bistatic OFDM-based ISAC system, including a description of the adopted OFDM frame structure. In the introduced system, incoming ISAC signals undergo synchronization and communication processing to reconstruct the originally transmitted OFDM frame. Afterwards, bistatic sensing is performed based on either pilot subcarriers only or on the full reconstructed transmitted OFDM frame. While the former approach yields limited radar sensing performance, the latter requires virtually error-free communication which is, e.g., achievable via channel coding, to avoid data decoding failures.
	\item The impact of data decoding failures on the resulting radar image is assessed and discussed. To mitigate this effect, a modulation symbol scrambling technique after channel coding is proposed.
	\item A detailed investigation of the effects of residual synchronization mismatches between transmitter and receiver on communication and radar is performed. The presented investigations cover time, frequency, and sampling frequnecy offset (TO, FO, SFO) in terms of error vector magnitude (EVM) as well as signal-to-noise ratio (SNR) and addional effects in the radar image.
	\item A measurement-based verification of the designed bistatic OFDM-based ISAC system is performed at \SI{79}{GHz}.
	
\end{itemize}

The remainder of this article is organized as follows. Sec.~\ref{sec:SystemModell} presents a thorough description of the designed bistatic OFDM-based ISAC system including the OFDM frame structure, the OTA synchronization, and the reconstruction of the transmitted OFDM frame. Based on that, Sec.~\ref{sec:SystemDesignAnalysis} analyzes the effects of data decoding failure and synchronization mismatches. Next, Sec.~\ref{sec:DemonstrationsmmWaveFrequencies} presents the measurement-based verification at \SI{79}{GHz}. Finally, concluding remarks are given in Sec.~\ref{sec:Conclusion}.

\section{System Model}\label{sec:SystemModell}
In the considered bistatic single-input single-output (SISO) OFDM-based ISAC system, a non-collocated transmitter-receiver pair performs radar sensing and communication through the same link. In this context, Fig.~\ref{figure:OFDMFramev2} shows the structure of the transmitted OFDM frame \mbox{$\mathbf{X}\in\mathbb{C}^{N\times M}$} in the discrete-frequency domain. The frame contains \mbox{$M\in\mathbb{N}_{>0}$} OFDM symbols, each consisting of \mbox{$N\in\mathbb{N}_{>0}$} subcarriers, where \mbox{$n \in \{0, 1, \dots , N-1\}$} and \mbox{$m \in \{0, 1, \dots, M-1\}$} are the subcarrier and OFDM symbol index. The $M$ transmitted OFDM symbols are composed of \mbox{$M_{\mathrm{ps}}\in\mathbb{N}_{>0}$} preamble symbols for the OTA synchronization and \mbox{$M_{\mathrm{pl}}\in\mathbb{N}_{>0}$} payload symbols. Within the payload symbols, pilots are inserted for equalization, where $M_{\mathrm{t}}\in\mathbb{N}_{>0}$ and $N_{\mathrm{f}}\in\mathbb{N}_{>0}$ describe the pilot spacing between OFDM symbols and subcarriers \cite{PaperSit}. As the transmitted modulation symbols must be known at the receiver for OFDM radar processing, pilots can be used to obtain a first radar image. However, this leads to a trade-off between radar and communications in terms of their optimal functionality \cite{unserPaper}. For the radar application, a large number of pilots is advantageous, as the processing gain increases. In contrast, it is worthwhile for the communication system to reduce the number of pilots to a minimum, as each pilot occupies space for potential payload data. This trade-off can be avoided in case of error-free communication, as it allows the receiver to reconstruct the transmitted OFDM frame. Consequently, the full OFDM frame is known to the receiver and the number of pilots can be minimized according to the maximum expected propagation delay $\tau_{\mathrm{exp}}$ and Doppler shift $f_{\mathrm{D,exp}}$. In this case, the data rate is maximized while taking advantage of the full processing gain. As discussed in \cite{PaperSit}, $M_{\mathrm{t}}$ and $N_{\mathrm{f}}$ can be calculated as
\begin{align}
M_{\mathrm{t}}&\leq \dfrac{1}{2\cdot f_{\mathrm{D,exp}}\cdot T_{\mathrm{OFDM}}} \label{Mt}
\end{align}
and
\begin{align}
N_{\mathrm{f}}&\leq \dfrac{1}{2\cdot \Delta f\cdot \tau_{\mathrm{exp}}} \label{Nf}\; ,
\end{align}
where $T_{\mathrm{OFDM}}$ is the duration of the OFDM symbol with cyclic prefix (CP) and $\Delta f$ is the subcarrier spacing.
\begin{figure}[t!]
	\centering
	\begin{tikzpicture}[scale=0.38]
		\begin{scope}
			\draw[ultra thick, step=1, draw=black!40, fill=3] (0,0) grid (18,10) rectangle (0,0);
			{}
			
			\draw[ultra thick, step=1, draw=black!40, fill=1] (0,0) grid (2,10) rectangle (0,0);
			
			\draw[ultra thick, step=1, draw=black!40, fill=2] (2,0) grid (3,1) rectangle (2,0);
			\draw[ultra thick, step=1, draw=black!40, fill=2] (2,3) grid (3,4) rectangle (2,3);
			\draw[ultra thick, step=1, draw=black!40, fill=2] (2,6) grid (3,7) rectangle (2,6);
			\draw[ultra thick, step=1, draw=black!40, fill=2] (2,9) grid (3,10) rectangle (2,9);
			
			\draw[ultra thick, step=1, draw=black!40, fill=2] (7,0) grid (8,1) rectangle (7,0);
			\draw[ultra thick, step=1, draw=black!40, fill=2] (7,3) grid (8,4) rectangle (7,3);
			\draw[ultra thick, step=1, draw=black!40, fill=2] (7,6) grid (8,7) rectangle (7,6);
			\draw[ultra thick, step=1, draw=black!40, fill=2] (7,9) grid (8,10) rectangle (7,9);
			
			\draw[ultra thick, step=1, draw=black!40, fill=2] (12,0) grid (13,1) rectangle (12,0);
			\draw[ultra thick, step=1, draw=black!40, fill=2] (12,3) grid (13,4) rectangle (12,3);
			\draw[ultra thick, step=1, draw=black!40, fill=2] (12,6) grid (13,7) rectangle (12,6);
			\draw[ultra thick, step=1, draw=black!40, fill=2] (12,9) grid (13,10) rectangle (12,9);
			
			\draw[ultra thick, step=1, draw=black!40, fill=2] (17,0) grid (18,1) rectangle (17,0);
			\draw[ultra thick, step=1, draw=black!40, fill=2] (17,3) grid (18,4) rectangle (17,3);
			\draw[ultra thick, step=1, draw=black!40, fill=2] (17,6) grid (18,7) rectangle (17,6);
			\draw[ultra thick, step=1, draw=black!40, fill=2] (17,9) grid (18,10) rectangle (17,9);
			
			\draw[ultra thick, step=1, draw=black!40, fill=1] (1,-3) grid (2,-2) rectangle (1,-3);
			\node[right] at (2,-2.6) {Preamble};
			\draw[ultra thick, step=1, draw=black!40, fill=2] (7,-3) grid (8,-2) rectangle (7,-3);
			\node[right] at (8,-2.6) {Pilots};
			\draw[ultra thick, step=1, draw=black!40, fill=3] (13,-3) grid (14,-2) rectangle (13,-3);
			\node[right] at (14,-2.6) {Payload};

			\draw [stealth-stealth,black,line width=0.5mm](0,12) -- (0,0) node [rotate=90,pos=0.5,above] {Subcarrier $n = 0, 1, ... , N-1$} -- (20,0) node[pos=0.5,below] {OFDM symbol $m = 0, 1, ... , M-1$};
			
			\draw [very thick,|-|](7.5,10.5) -- (12.5,10.5) node[pos=0.5,above] {$M_{\mathrm{t}}$};
			\draw [very thick,|-|](18.5,6.5) -- (18.5,9.5) node[pos=0.5,right] {$N_{\mathrm{f}}$};
			
		\end{scope}
	\end{tikzpicture}
	\caption{Structure of the transmitted OFDM frame \textbf{X}}
	\label{figure:OFDMFramev2}
\end{figure}
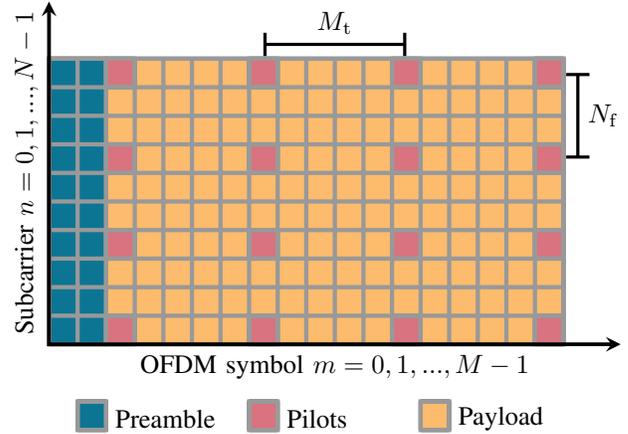
\begin{figure*}[!b]
	\hrulefill
	\vspace*{4pt}
	\setcounter{equation}{4}
	\begin{small}
		\begin{align}\label{YSFO}
		(\mathbf{Y}^{\mathrm{SFO}})_{n,m}&= (\mathbf{\tilde{Y}})_{n,m}\underbrace{\cdot\dfrac{\sin(\pi \delta_{\mathrm{off}}n)}{N\sin(\frac{\pi \delta_{\mathrm{off}}n}{N})}}_{\text{attenuation of the amplitude along the subcarriers}}\cdot\underbrace{\vphantom{ \dfrac{\sin(\pi \delta_{\mathrm{off}}n)}{N\sin(\frac{\pi \delta_{\mathrm{off}}n}{N})} }e^{\mathrm{j}2\pi \frac{m(N+N_{\mathrm{CP}})+N_{\mathrm{CP}}}{N}\delta_{\mathrm{off}}n}\cdot e^{\mathrm{j}\pi \frac{N-1}{N}\delta_{\mathrm{off}}n}}_{\text{time-varying phase shift}}\notag\\
		&+ \underbrace{\sum_{n'=0,n'\neq n}^{N-1} (\mathbf{\tilde{Y}})_{n,m}\cdot \dfrac{\sin(\pi ((1+\delta_{\mathrm{off}})n'-n))}{N\sin(\frac{\pi ((1+\delta_{\mathrm{off}})n'-n)}{N})}\cdot e^{\mathrm{j}2\pi \frac{m(N+N_{\mathrm{CP}})+N_{\mathrm{CP}}}{N}\delta_{\mathrm{off}}n'}\cdot e^{\mathrm{j}\pi \frac{N-1}{N}((1+\delta_{\mathrm{off}})n'-n)}}_{\text{ICI}}
		\end{align}
	\end{small}
\end{figure*}\setcounter{equation}{2}To enable forward error correction (FEC), the transmitted bits are passed through a low-density parity-check (LDPC) encoder for 5G New Radio (NR) \cite{Proakis,Tahir,Richardson}. Afterwards, the coded bits undergo a quadrature phase-shift keying (QPSK) modulation with Gray coding and together with pilots and preamble symbols, they are placed into a frame of bandwidth $B$, as in Fig.~\ref{figure:OFDMFramev2}. The first part of the preamble consits of two OFDM symbols for time and frequency synchronization according to the algorithm proposed by Schmidl and Cox (S\&C) in \cite{PaperSuC}. The second part includes additional OFDM symbols for sampling frequency synchronization according to the Tsai algorithm \cite{PaperTsai}. Next, each OFDM symbol is transformed into the discrete-time domain by an inverse discrete Fourier transform (IDFT). After a CP of length $N_{\mathrm{CP}}\in\mathbb{N}_{>0}$ is prepended to each OFDM symbol, the frame is converted into the serial transmit sequence and the real and imaginary parts undergo digital-to-analog (D/A) conversion with sampling frequency $f_{\mathrm{s}} \geq B$. The resulting continuous-time domain baseband signal $x(t)\in\mathbb{C}$ is low-pass filtered, upconverted to the carrier frequency $f_{\mathrm{c}} \gg B$ by an I/Q mixer, and finally radiated by the transmit antenna.

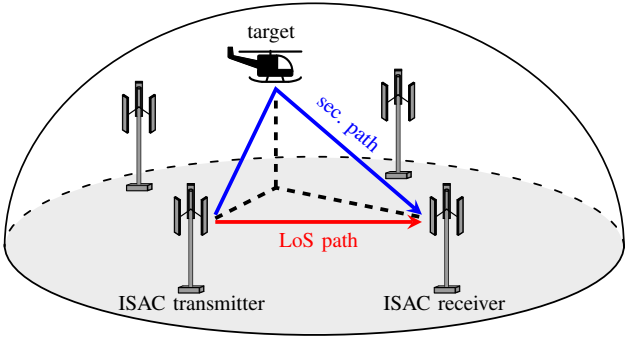
\begin{figure}[t!]
	\centering
	\begin{tikzpicture}[scale=0.175]
	\node[] at (0,0) {\includegraphics[scale=0.4]{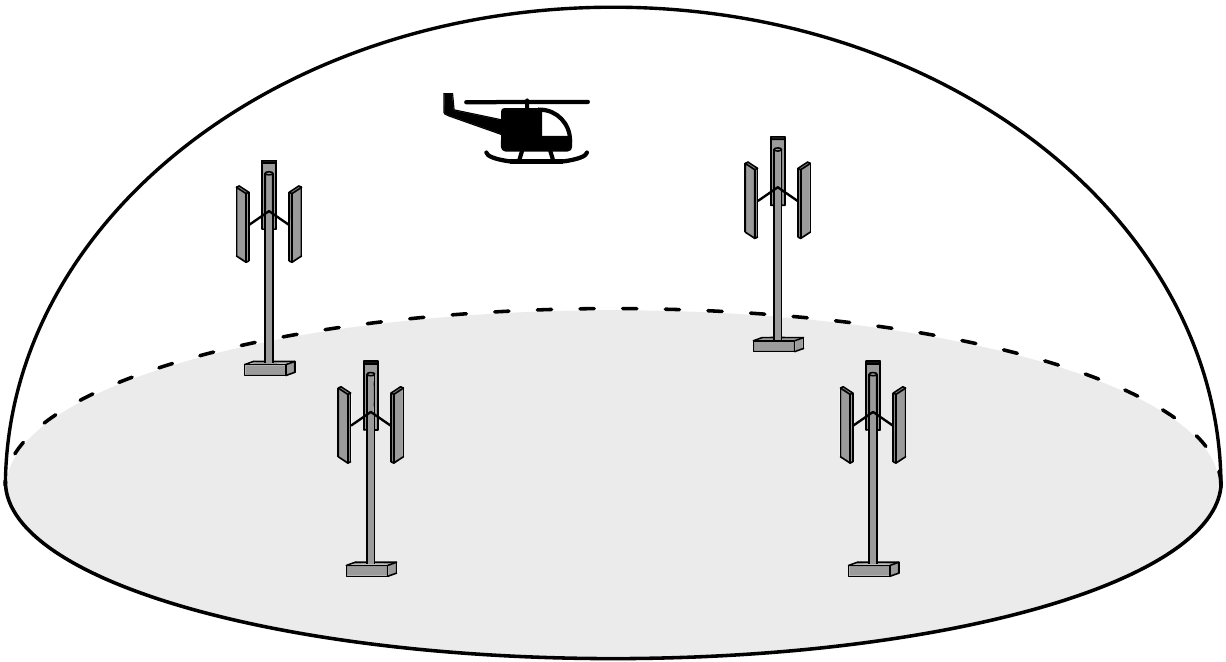}};
	\begin{scope}
	\draw [-stealth,red,line width=0.5mm](-7.5,-4.1) -- (8.2,-4.1) node[pos=0.5,below] {\footnotesize LoS path};
	\draw [-stealth,blue,line width=0.5mm](-7.5,-3.55) -- (-2.9,6) -- (8.2,-3.55) node[pos=0.4,above,rotate=-41.5] {\footnotesize sec. path};
	\draw [-,black,dashed,line width=0.5mm](-7.5,-3.825) -- (-2.9,-1.5) -- (8.2,-3.75) node[pos=0.5,right] {};
	\draw [-,black,dashed,line width=0.5mm](-2.9,-1.5) -- (-2.9,5.5) node[pos=0.5,right] {};
	
	\end{scope}
	
	\node [below,] at (-3.25,11.5) {\footnotesize target};
	\node [below] at (-9.3,-9) {\footnotesize ISAC transmitter};
	\node [below] at (9.9,-9) {\footnotesize ISAC receiver};
	
	\end{tikzpicture}
	\caption[]{Bistatic scenario with one target}
	\label{figure:Basestations2}
\end{figure}
As stated in \cite{unserPaper} and illustrated for a single target in Fig.~\ref{figure:Basestations2}, the noiseless analog continuous-time domain baseband receive signal $\tilde{y}(t)\in\mathbb{C}$ after propagation, analog receiving processing, and downconversion can be written as
\begin{align}
\tilde{y}(t)&= \alpha_{\mathrm{0}} \cdot x(t-\tau_{\mathrm{0}}-\tau_{\mathrm{\Delta}}) \cdot e^{\mathrm{j}2\pi f_{\mathrm{D,0}}t} \cdot e^{\mathrm{j}(2\pi f_{\mathrm{c,off}}t+\phi_{\mathrm{c,off}})} \notag\\
&+ \sum_{h=1}^{H-1} \alpha_{h} \cdot x(t-\tau_{h}-\tau_{\mathrm{\Delta}}) \cdot e^{\mathrm{j}2\pi f_{\mathrm{D,}h}t} \cdot e^{\mathrm{j}(2\pi f_{\mathrm{c,off}}t+\phi_{\mathrm{c,off}})}. \label{yttilde}
\end{align}
As can be seen in \eqref{yttilde}, the transmitted signal propagates through $H\in\mathbb{N}_{>0}$ paths, each path being specified by the index $h \in \{0, 1, \dots, H-1\}$. These can be divided into a main path ($h=0$), which is assumed to be a line-of-sight (LoS) path, and several secondary paths ($h=1,...,H-1$). While propagating through the different paths, the transmit signal experiences attenuation $\alpha_{h}$, delay $\tau_{h}$, and Doppler shift $f_{\mathrm{D,}h}$. Since transmitter and receiver are spatially separated and not synchronized, they refer to different time references, local oscillators (LO), and sampling clock references. As a result, the received signal experiences an additional symbol time offset (STO) $\tau_{\mathrm{\Delta}}$, carrier frequency offset (CFO) $f_{\mathrm{c,off}}$, and carrier phase offset (CPO) $\phi_{\mathrm{c,off}}$. In the next step, the received baseband signal undergoes analog-to-digital (A/D) conversion. Deviations in the sampling frequency between the digital-to-analog converters (DAC) in the transmitter and the analog-to-digital converters (ADC) in the receiver cause a sampling frequency offset (SFO) $\delta_{\mathrm{off}}$. Thus, the resulting discrete-time domain signal $y[\eta]\in\mathbb{C}$ can be written as
\begin{align}
y[\eta] = \tilde{y}(\eta T_{\mathrm{s}}(1+\delta_{\mathrm{off}}))\; ,\label{yeta}
\end{align}
where $T_{\mathrm{s}}$ is the sampling time \cite{unserPaper}. Following \cite{BuchZF}, and assuming an SFO small enough not to generate ISI, the effects of SFO on the received OFDM frame can be described by \eqref{YSFO}, where $(\mathbf{\tilde{Y}})_{n,m}$ is the synchronized OFDM frame and $n'$ is a second index along the subcarriers. As can be seen from \eqref{YSFO}, SFO leads to an attenuation of the amplitude along the subcarriers, a time-varying phase shift as well as inter-carrier interference (ICI).

\subsection{Synchronization}\label{sec:Synchronization}
Consequently, a sufficient synchronization is required for a bistatic OFDM-based ISAC system. As previously mentioned, an OTA synchronization is performed, which means that the synchronization is based on the preamble and pilot symbols passed on to the receiver with the transmit signal. The synchronization is performed with respect to the strongest path, which is assumed to be the LoS path as it has the shortest propagation time and does not suffer from reflections off scatterers. The receiver cannot distinguish between $\tau_{\mathrm{0}}$ and $\tau_{\mathrm{\Delta}}$, as well as between $f_{\mathrm{D,0}}$ and $f_{\mathrm{c,off}}$. Therefore, the time offset (TO) is defined as \mbox{$\tau_{\mathrm{off}} = \tau_{\mathrm{0}} + \tau_{\mathrm{\Delta}}$} and the frequency offset (FO) as \mbox{$f_{\mathrm{off}} = f_{\mathrm{D,0}} + f_{\mathrm{c,off}}$}. In addition, the aforementioned SFO arises. Before communication and radar processing, these three offsets must be corrected at the receiver.

First, the TO has to be compensated to determine a starting point of the OFDM signal. Using the first two preamble symbols, a coarse TO and FO is estimated based on the S\&C algorithm \cite{PaperSuC}. For fine TO estimation, a cross-correlation between the received signal $y[\eta]$ and the first preamble symbol is performed \cite{unserPaper,PaperSit}. To reduce the complexity of the correlation, $y[\eta]$ is shortened around the coarse starting point. Furthermore, the coarse FO is corrected for the shortened version of $y[\eta]$ to avoid increased correlation side lobes due to high frequency shifts. For this purpose, the discrete-time domain signal is multiplied with exp($-\mathrm{j2\pi}\hat{f}_{\mathrm{off}}\eta$), where $\hat{f}_{\mathrm{off}}$ is the coarse FO.

After finding the starting point of the received signal, the SFO is estimated based on the remaining preamble, which consists of several pairs of identical OFDM symbols. The design of these symbol pairs and the estimation of the SFO via weighted least-squares algorithm is performed according to \cite{PaperTsai}. For appropriate SFO estimation, the coarse FO must also be corrected for this second preamble part. To correct the SFO, the whole receive signal is interpolated by a multirate finite impulse response (FIR) filter \cite{gardner1993,erup1993}  and fed into a sample rate converter based on a polynomial filter \cite{farrow1988}. Finally, the signal is decimated to the original number of samples using a second multirate FIR filter. Due to inaccuracies, a residual SFO may still remain and is corrected with a delay time profile (DTP) based on the pilots. Therefore, the received signal is converted into a frame shape and transformed into the discrete-frequency domain by a discrete Fourier transform (DFT). Then, the pilots within this frame undergo elementwise division by the transmitted pilots and an IDFT with zero padding along the subcarrier direction. Thus, the final DTP shows the delay over the symbol index, whereby the delay accuracy is improved due to zero padding.

As can be seen from the time-varying phase shift term in \eqref{YSFO}, this delay changes approximately linearly with the symbol index in case of a residual SFO. This residual SFO can be estimated by determining the delays of a propagation path \cite{Wu2012} in the first and last pilot symbol $\tau_{\mathrm{begin}}$ and $\tau_{\mathrm{end}}$. Theoretically, this would be possible with any propagation path, since an SFO affects all paths equally. However, it is convenient to use the LoS path, as it has the highest power level compared to the other propagation paths. Finally, the residual SFO $\hat{\delta}_{\mathrm{off,res}}$ can be calculated in parts per million (ppm) with
\setcounter{equation}{5}
\begin{align}
\hat{\delta}_{\mathrm{off,res}}&= \dfrac{\tau_{\mathrm{end}}-\tau_{\mathrm{begin}}}{T_{\mathrm{s}}} \cdot \dfrac{10^{6}}{(N+N_{\mathrm{CP}})\cdot (m_{\mathrm{end}}-m_{\mathrm{begin}}+1)}, \label{RestSFOFormel}
\end{align}
where $m_{\mathrm{begin}}$ and $m_{\mathrm{end}}$ represent the index of the first and last OFDM symbol containing pilots. Since the estimates of $\tau_\mathrm{begin}$ and $\tau_\mathrm{end}$ are affected by noise and the quantized delay estimation enabled by the zero-padded IDFT, all delays estimated at OFDM symbols containing pilots can be used to obtain a more accurate SFO estimate by a least-squares estimator based on the same principle from \eqref{RestSFOFormel}. The correction of the residual SFO is performed with FIR and polynomial filters, as before.

In the last step, the FO is compensated. Therefore, the whole signal is corrected with the previous estimate of the coarse FO. In order to be able to determine a residual FO, a delay Doppler profile (DDP) is created based on the pilots. Out of that, a residual frequency shift of the LoS path can be obtained \cite{PaperSit}. Since all delays have already been corrected and the LoS path is at the known delay tap of \SI{0}{\second}, it is sufficient for the DDP to execute the corresponding FFT only over the first line of a DTP.

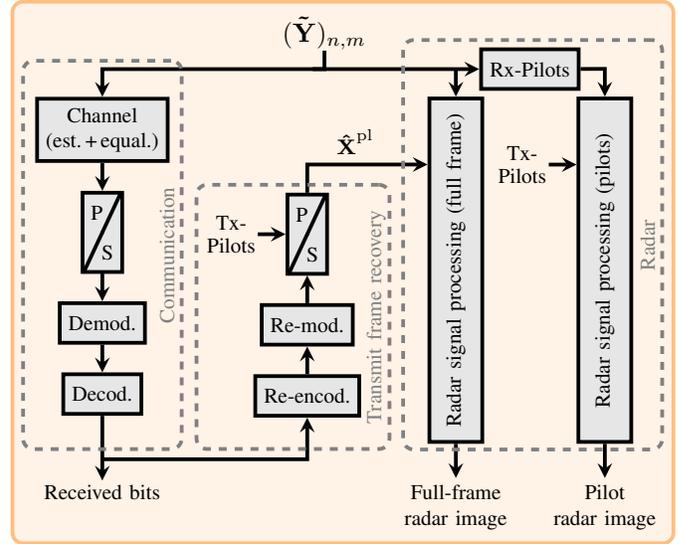
\begin{figure}[t!]
	\centering
	\begin{tikzpicture}[scale=0.35,rotate=0]
	
	\begin{scope}[]
	\node [rectangle,draw,text = black,fill = gray!20,line width=0.5mm,rotate=0, node distance=0em, text centered, minimum height=2em, minimum width=2em, yshift=-4.75em,align=center] (H) at (-26em,0em) {\footnotesize Channel\\[-0.5ex] \footnotesize (est.\,+\,equal.)};
	\draw [stealth-,black,line width=0.5mm](H.north) -- ++(0em,3em) -- ++(24em,0em) -- ++(0em,2em) node[pos=0.5,above]() {$(\mathbf{\tilde{Y}})_{n,m}$};
	
	\node [rectangle,draw,text = black,fill = gray!20,line width=0.5mm,rotate=0, node distance=0em, text centered, minimum height=3em, minimum width=1.5em, below of=H,xshift=0em,yshift=-3.88em] (PS1) {};
	\draw [-,black,line width=0.5mm,line cap=round](PS1.center) -- ++(\SPup) node[right]() {};
	\draw [-,black,line width=0.5mm,line cap=round](PS1.center) -- ++(\SPdown) node[right]() {};
	\node [black,rotate=0,xshift=-0.2em,yshift=0.75em] at (PS1.center) {\footnotesize P};
	\node [black,rotate=0,xshift=0.2em,yshift=-0.75em] at (PS1.center) {\footnotesize S};
	\draw [stealth-,black,line width=0.5mm](PS1.north) -- ++(0em,3em);
	
	\node [rectangle,draw,text = black,fill = gray!20,line width=0.5mm,rotate=0, below of=PS1, node distance=0em, text centered, minimum height=1.5em, minimum width=2em, yshift=-3.43em,align=center] (Demod) {\footnotesize Demod.};
	\draw [stealth-,black,line width=0.5mm](Demod.north) -- ++(0em,3em);
	
	\node [rectangle,draw,text = black,fill = gray!20,line width=0.5mm,rotate=0, below of=Demod, node distance=0em, text centered, minimum height=1.5em, minimum width=2em, yshift=-2.69em,align=center] (Decod) {\footnotesize Decod.};
	\draw [stealth-,black,line width=0.5mm](Decod.north) -- ++(0em,3em);
	
	\draw [-stealth,black,line width=0.5mm](Decod.south) -- ++(0em,-7em) node[pos=0.9,below,rotate=0]() {\footnotesize Received bits};
	\end{scope}
	
	\begin{scope}[xshift=20.2em,yshift=-7em]
	\node [rectangle,draw,text = black,fill = gray!20,line width=0.5mm,rotate=0, node distance=0em, text centered, minimum height=1.5em, minimum width=2em, yshift=0em,align=center] (DezPiloten) at(0,0) {\footnotesize Rx-Pilots};
	\draw [stealth-,black,line width=0.5mm](DezPiloten.west) -- ++(-18em,0em);
	\end{scope}
	
	\begin{scope}[xshift=4.3em,yshift=-0.3em]
	\node [rectangle,draw,text = black,fill = gray!20,line width=0.5mm,rotate=90, node distance=0em, text centered, minimum height=2em, minimum width=13em, yshift=0em,align=center] (RadarPiloten) at (24em,-28.45em) {\footnotesize Radar signal processing (pilots)};
	\draw [stealth-,black,line width=0.5mm](RadarPiloten.east) -- ++(0em,3em) -- (DezPiloten);
	\draw [stealth-,black,line width=0.5mm,align=center](RadarPiloten.15) -- ++(-3em,0em) node[pos=0.7,left,rotate=0](Piloten) {\footnotesize Tx-\\[-0.75ex]\footnotesize Pilots};
	
	\node [rectangle,draw,text = black,fill = gray!20,line width=0.5mm,rotate=90, node distance=0em, text centered, minimum height=2em, minimum width=13em, yshift=0em,align=center] (RadarFF) at (7.9em,-28.45em) {\footnotesize Radar signal processing (full frame)};
	\draw [stealth-,black,line width=0.5mm](RadarFF.east) -- ++(0em,3em);
	\end{scope}
	
	\begin{scope}[]
	\draw [-stealth,black,line width=0.5mm,align=center](RadarFF.west) -- ++(0em,-3.7em) node[pos=0.9,below,rotate=0]() {\footnotesize Full-frame\\[-0.5ex] \footnotesize radar image};
	\draw [-stealth,black,line width=0.5mm,align=center](RadarPiloten.west) -- ++(0em,-3.7em) node[pos=0.9,below,rotate=0]() {\footnotesize Pilot\\[-0.5ex] \footnotesize radar image};
	\end{scope}
	
	\begin{scope}[]	
	\node [rectangle,draw,text = black,fill = gray!20,line width=0.5mm,rotate=0, node distance=0em, text centered, minimum height=3em, minimum width=1.5em, below of=H,xshift=7.75em,yshift=-3.95em] (PS2) {};
	\draw [-,black,line width=0.5mm,line cap=round](PS2.center) -- ++(\SPup) node[right]() {};
	\draw [-,black,line width=0.5mm,line cap=round](PS2.center) -- ++(\SPdown) node[right]() {};
	\node [black,rotate=0,xshift=-0.2em,yshift=0.75em] at (PS2.center) {\footnotesize P};
	\node [black,rotate=0,xshift=0.2em,yshift=-0.75em] at (PS2.center) {\footnotesize S};
	\draw [-stealth,black,line width=0.5mm](PS2.north) -- ++(0em,3.05em) -- ++(13em,0em) node[pos=0.4,above,rotate=0]() {\footnotesize $\mathbf{\hat{X}}^{\mathrm{pl}}$};
	
	\node [rectangle,draw,text = black,fill = gray!20,line width=0.5mm,rotate=0, below of=PS2, node distance=0em, text centered, minimum height=1.5em, minimum width=2em, yshift=-3.43em,align=center] (Mod) {\footnotesize Re-mod.};
	\draw [-stealth,black,line width=0.5mm](Mod.north) -- ++(0em,3em);
	
	\node [rectangle,draw,text = black,fill = gray!20,line width=0.5mm,rotate=0, below of=Mod, node distance=0em, text centered, minimum height=1.5em, minimum width=2em, yshift=-2.69em,align=center] (Reencod) {\footnotesize Re-encod.};
	\draw [-stealth,black,line width=0.5mm](Reencod.north) -- ++(0em,3em);
	
	\draw [stealth-,black,line width=0.5mm,align=center](PS2.west) -- ++(-3em,0em) node[pos=0.7,left,rotate=0](Piloten) {\footnotesize Tx-\\[-0.75ex]\footnotesize Pilots};
	\draw [stealth-,black,line width=0.5mm](Reencod.south) -- ++(0em,-4.5em) -- ++(-22em,0em);
	\end{scope}
	
	\begin{scope}[xshift=-26em,yshift=-27.1em]
	\node [rectangle,rounded corners,draw,gray,dashed,line width=0.5mm,rotate=0, node distance=0em, text centered, minimum height=14.7em, minimum width=6em] (Komm) at(0,0) {};
	\node[gray,rotate=90,above of=Komm,xshift=0em,yshift=-5.3em] () {\footnotesize Communication};
	\end{scope}
	
	\begin{scope}[xshift=-5.4em,yshift=-33.8em]
	\node [rectangle,rounded corners,draw,gray,dashed,line width=0.5mm,rotate=0, node distance=0em, text centered, minimum height=10em, minimum width=7.3em] (Rek) at(0,0) {};
	\node[gray,rotate=90,above of=Rek,xshift=0em,yshift=-6em] () {\footnotesize Transmit frame recovery};
	\end{scope}
	
	\begin{scope}[xshift=20.5em,yshift=-26em]
	\node [rectangle,rounded corners,draw,gray,dashed,line width=0.5mm,rotate=0, node distance=0em, text centered, minimum height=15.5em, minimum width=9.8em] (Radar) at(0,0) {};
	\node[gray,rotate=90,above of=Radar,xshift=0em,yshift=-7.2em] () {\footnotesize Radar};
	\end{scope}	
	
	\begin{scope}[on background layer]
	\node [rectangle,rounded corners,draw,orange!50,fill=orange!10,line width=0.5mm,rotate=0, node distance=0em, text centered, minimum height=20.5em, minimum width=25em,xshift=0em,yshift=0em] () at(0,-29em) {};
	\end{scope}
	
	\end{tikzpicture}
	\caption{Block diagram of the communication and radar signal processing at the receiver}
	\label{figure:OFDMBlockschaltbildRx}
\end{figure}

After successful synchronization, the preamble symbols are discarded and the discrete-time serial received signal is transfered into its parallel frame shape. Next, the CP is removed and the frame is converted into the discrete-frequency domain by a DFT. Consequently, and following \cite{Smaini}, the received and synchronized OFDM frame can be expressed as
\begin{align}
(\mathbf{\tilde{Y}})_{n,m}&= \alpha_{\mathrm{0}} \cdot (\mathbf{X})_{n,m} \cdot e^{\mathrm{j}\phi_{\mathrm{c,off}}} + \sum_{h=1}^{H-1} \alpha_{h} \cdot (\mathbf{X})_{n,m}\notag\\
&\cdot e^{-\mathrm{j}2\pi n\Delta f(\tau_{h}-\tau_{\mathrm{0}})}\cdot e^{\mathrm{j}2\pi (f_{\mathrm{D,}h}-f_{\mathrm{D,0}})mT_{\mathrm{OFDM}}}\cdot e^{\mathrm{j}\phi_{\mathrm{c,off}}}. \label{Ysynch}
\end{align}
This frame will be further processed to obtain the communication and radar information. Fig.~\ref{figure:OFDMBlockschaltbildRx} shows the block diagram of the communication and radar signal processing at the receiver described in the following.

\subsection{Communication Processing}\label{sec:CommunicationProcessing}
The first step of communication processing is calculating a channel estimate by dividing the received pilots by the transmitted pilots \cite{Coleri,Li}. In order to adapt this channel estimate to the size of the full OFDM frame, a spline interpolation in subcarrier and symbol direction is performed \cite{PaperSit}. Subsequently, the received OFDM frame in \eqref{Ysynch} is divided by the interpolated channel estimate to perform equalization and the frame is transformed into a serial modulation data sequence. The following QPSK demodulation and soft decision LDPC decoding, which removes bit errors in the received signal, complete the communication processing and ideally lead to the succesful recovery of the transmitted bits.

\subsection{Radar Processing}\label{sec:RadarProcessing}
Due to the non-collocated transmitter and receiver, the receiver has no information about the whole transmitted OFDM frame, which, however, is required to perform OFDM radar signal processing. Therefore, the recovered bit sequence from communication forms the basis for a radar image with the full OFDM frame utilizing the total processing gain. For this purpose, the received bit sequence is re-encoded and modulated again. Converting into the frame shape and inserting the pilots leads to a recovery of the transmitted OFDM payload frame $\mathbf{\hat{X}}^{\mathrm{pl}}$. Afterwards, the usual OFDM radar processing is performed, where the received OFDM frame in the discrete-frequency domain undergoes an elementwise division by the recovered transmit frame, as well as an IDFT along the subcarrier direction and a DFT along the symbol direction \cite{Giroto}. In addition, a Blackman window is applied in both subcarrier and OFDM symbol direction to suppress sidelobes, as well as zero-padding for improved accuracy. As can be seen from \eqref{Ysynch}, the measured distance and Doppler shift shown in the radar image are based on the difference in travel time and Doppler shift between a secondary path and the LoS path. Because the full OFDM frame is used here, the obtained radar image is called full-frame radar image in the following. However, to avoid artifacts in the obtained full-frame radar image due to incorrect estimation of the transmitted communication bits in the payload, error-free communication is required.

Regardless of the aforementioned frame reconstruction, the received OFDM frame can be limited to the pilots to perform the radar processing only based on pilots. In contrast to the full-frame radar image, this pilot radar image is not dependent on the communication and can be generated even in case of bit errors. However, the processing gain is reduced, as well as the unambiguous range and Doppler shift. Following \cite{unserPaper}, the radar performance parameters of the bistatic OFDM-based ISAC system are listed in Tab.~\ref{tab:radarParameters}.

\begin{table}[!t]
	\renewcommand{\arraystretch}{1.5}
	\arrayrulecolor[HTML]{000000}
	\setlength{\arrayrulewidth}{.1mm}
	\setlength{\tabcolsep}{4pt}
	
	\centering
	\captionsetup{width=43pc,justification=centering,labelsep=newline}
	\caption{Radar performance parameters of the bistatic OFDM-based ISAC system}
	\begin{tabular}{|cc|}
		\hhline{|==|}
		\multicolumn{1}{|c|}{\textbf{Processing gain}}      & $G_\mathrm{p} = \lceil N/N_\mathrm{f}\rceil\cdot \lceil M_\mathrm{pl}/M_\mathrm{t}\rceil$ \\ \hline
		\multicolumn{1}{|c|}{\textbf{Range resolution}}     & $\Delta R = c_0/B$ \\ \hline
		\multicolumn{1}{|c|}{\textbf{Max. unamb. range}}    & $R_\mathrm{max,ua} = c_0/B\cdot\lceil N/N_\mathrm{f}\rceil$ \\ \hline
		\multicolumn{1}{|c|}{\textbf{Max. ISI-free range}}    & $R_\mathrm{max,ISI} = c_0/B\cdot N_\mathrm{CP}$ \\ \hline
		\multicolumn{1}{|c|}{\textbf{Doppler shift resolution}}  & $\Delta f_\mathrm{D} = B/\left[\left(N+N_\mathrm{CP}\right)M_\mathrm{pl}\right]$ \\ \hline
		\multicolumn{1}{|c|}{\textbf{Max. unamb. Doppler shift}} & $f_\mathrm{D,max,ua} = \pm B/\left[2M_\mathrm{t}\cdot \left(N+N_\mathrm{CP}\right)\right]$ \\ \hline
		\multicolumn{1}{|c|}{\textbf{Max. ICI-free Doppler shift}} & $f_\mathrm{D,max,ICI} = \pm B/(10N)$ \\ \hhline{|==|}
	\end{tabular}
	\label{tab:radarParameters}
\end{table}

\section{System Design and Analysis}\label{sec:SystemDesignAnalysis}
Considering the previously introduced system model, there are two key points to ensure proper operation of the bistatic OFDM-based ISAC system: reconstruction of the transmitted OFDM frame and synchronization. Therefore, this section addresses the effects of an incorrectly reconstructed OFDM frame as well as the impact of remaining synchronization offsets. The following investigations are performed based on the bistatic OFDM-based ISAC system described in Sec.~\ref{sec:SystemModell}. For the subsequent simulations, a scenario with one LoS path between transmitter and receiver and another secondary path raised by a moving radar target is adopted. The corresponding OFDM signal, communication and radar parameters are listed in Tab.~\ref{tab:resultsParameters}.

\begin{table}[t!]
	\renewcommand{\arraystretch}{1.5}
	\arrayrulecolor[HTML]{000000}
	\setlength{\arrayrulewidth}{.1mm}
	\setlength{\tabcolsep}{4pt}
	
	\centering
	\captionsetup{width=43pc,justification=centering,labelsep=newline}
	\caption{OFDM signal, communication and radar parameters}
	\resizebox{\columnwidth}{!}{
		\begin{tabular}{|cc|c|}
			\hhline{|===|}
			\rowcolor{lightgray}
			\multicolumn{3}{|c|}{\textbf{OFDM signal parameters}} \\ \hhline{|===|}
			\multicolumn{1}{|c|}{\textbf{SNR LoS path}}      & \multicolumn{2}{c|}{$\mathit{SNR}_{\mathrm{LoS}}=\SI{15}{dB}$} \\ \hline
			\multicolumn{1}{|c|}{\textbf{SNR secondary path}}      & \multicolumn{2}{c|}{$\mathit{SNR}_{\mathrm{SP}}=\SI{-20}{dB}$} \\ \hline
			\multicolumn{1}{|c|}{\textbf{Bandwidth}}      & \multicolumn{2}{c|}{$B=\SI{1}{\giga \hertz}$} \\ \hline
			\multicolumn{1}{|c|}{\textbf{Sampling frequency}}      & \multicolumn{2}{c|}{$f_{\mathrm{s}}=\SI{2}{\giga \hertz}$} \\ \hline
			\multicolumn{1}{|c|}{\textbf{No. of subcarriers}}      & \multicolumn{2}{c|}{$N=2048$} \\ \hline
			\multicolumn{1}{|c|}{\textbf{Subcarrier spacing}}      & \multicolumn{2}{c|}{$\Delta f=\SI{488.28}{\kilo \hertz}$} \\ \hline
			\multicolumn{1}{|c|}{\textbf{CP length}}      & \multicolumn{2}{c|}{$N_\mathrm{CP}=512$} \\ \hline
			\multicolumn{1}{|c|}{\textbf{No. of preamble symbols}}      & \multicolumn{2}{c|}{$M_\mathrm{ps}=2+8$} \\ \hline
			\multicolumn{1}{|c|}{\textbf{No. of payload symbols}}      & \multicolumn{2}{c|}{$M_\mathrm{pl}=512$} \\ \hline
			\multicolumn{1}{|c|}{\textbf{Pilot spacing}}      & \multicolumn{2}{c|}{$N_\mathrm{f}=2$, $M_\mathrm{t}=7$} \\ \hhline{|===|}
			\rowcolor{lightgray}
			\multicolumn{3}{|c|}{\textbf{Communication performance parameters}} \\ \hhline{|===|}
			\multicolumn{1}{|c|}{\textbf{Channel coding and code rate}}      & \multicolumn{2}{c|}{LDPC, $r=2/3$} \\ \hline
			\multicolumn{1}{|c|}{\textbf{Data rate (information bits)}}      & \multicolumn{2}{c|}{$\mathcal{R}_\mathrm{comm}=\SI{0.97}{Gbit/s}$} \\ \hhline{|===|}
			\rowcolor{lightgray}
			\multicolumn{3}{|c|}{\textbf{Radar performance parameters}} \\ \hhline{|===|}
			\multicolumn{1}{|c|}{\multirow{2}{*}{\textbf{Processing gain}}}  & \multicolumn{2}{c|}{$G_\mathrm{p}=\SI{48.80}{dB}$ (pilots only)} \\
			\multicolumn{1}{|c|}{} & \multicolumn{2}{c|}{$G_\mathrm{p}=\SI{60.21}{dB}$ (full-frame)} \\ \hline
			\multicolumn{1}{|c|}{\textbf{Range resolution}}     & \multicolumn{2}{c|}{$\Delta R=\SI{0.30}{\meter}$} \\ \hline
			\multicolumn{1}{|c|}{\multirow{2}{*}{\textbf{Max. unamb. range}}}  & \multicolumn{2}{c|}{$R_\mathrm{max,ua}=\SI{306.99}{\meter}$ (pilots only)} \\
			\multicolumn{1}{|c|}{} & \multicolumn{2}{c|}{$R_\mathrm{max,ua}=\SI{613.97}{\meter}$ (full-frame)} \\ \hline
			\multicolumn{1}{|c|}{\textbf{Max. ISI-free range}}    & \multicolumn{2}{c|}{$R_\mathrm{max,ISI}=\SI{153.49}{\meter}$} \\ \hline
			\multicolumn{1}{|c|}{\textbf{Doppler shift resolution}}  & \multicolumn{2}{c|}{$\Delta f_\mathrm{D}=\SI{762.94}{\hertz}$} \\ \hline
			\multicolumn{1}{|c|}{\multirow{2}{*}{\textbf{Max. unamb. Doppler shift}}}  & \multicolumn{2}{c|}{$f_\mathrm{D,max,ua}=\SI{27.90}{\kilo\hertz}$ (pilots only)} \\
			\multicolumn{1}{|c|}{} & \multicolumn{2}{c|}{$f_\mathrm{D,max,ua}=\SI{195.31}{\kilo\hertz}$ (full  frame)} \\ \hline
			\multicolumn{1}{|c|}{\textbf{Max. ICI-free Doppler shift}} & \multicolumn{2}{c|}{$f_\mathrm{D,max,ICI}=\SI{48.83}{\kilo\hertz}$} \\ \hhline{|===|}
		\end{tabular}
	}
	\label{tab:resultsParameters}
\end{table}

\subsection{Impact of Data Decoding Failures}
\begin{figure}[t!]
	\centering
	\begin{subfigure}[t]{0.218\textwidth}
		\centering
		\psfrag{SER}[c][c]{Absolute bit errors}
		\includegraphics[width=\textwidth]{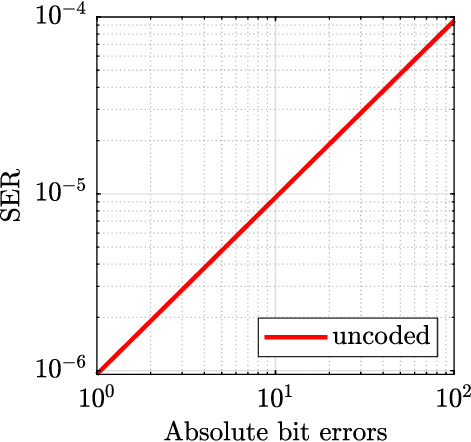}
		\caption{$M_{\mathrm{pl}}=512$, QPSK}
		\label{fig:falModsymbuncodiert}
	\end{subfigure}
	\hfill
	\begin{subfigure}[t]{0.218\textwidth}
		\centering
		\includegraphics[trim = 0.85cm 0.08cm 1.75cm 0.3cm,clip,width=\textwidth]{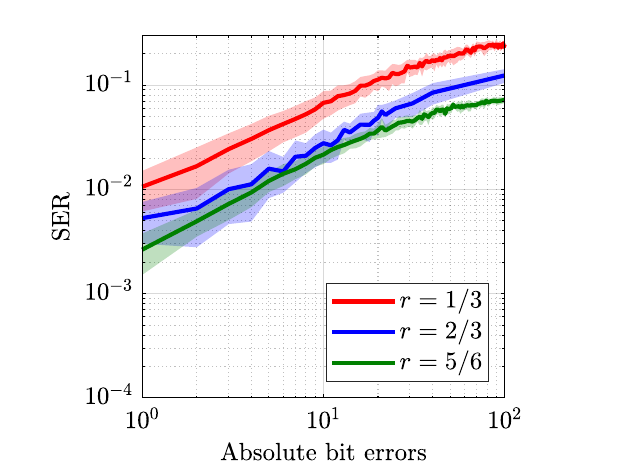}
		\caption{$M_{\mathrm{pl}}=512$, QPSK}
		\label{fig:falModsymbCoderate}
	\end{subfigure}
	\par\bigskip
	\begin{subfigure}{0.218\textwidth}
		\includegraphics[trim = 0.85cm 0.08cm 1.75cm 0.3cm,clip,width=\textwidth]{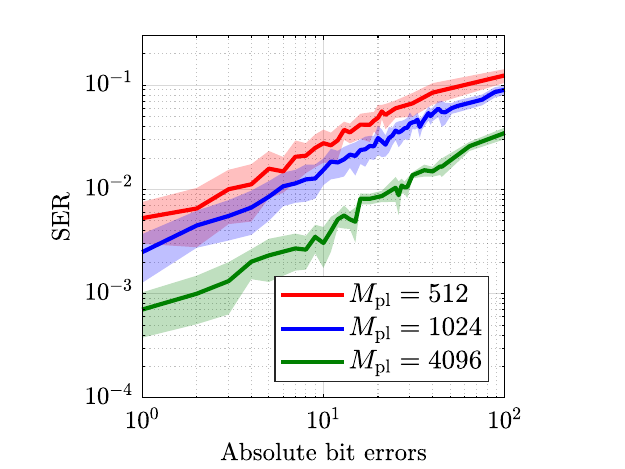}
		\caption{$r=2/3$, QPSK}
		\label{fig:falModsymbFrame}
	\end{subfigure}
	\hfill
	\begin{subfigure}{0.218\textwidth}
		\includegraphics[trim = 0.85cm 0.08cm 1.75cm 0.3cm,clip,width=\textwidth]{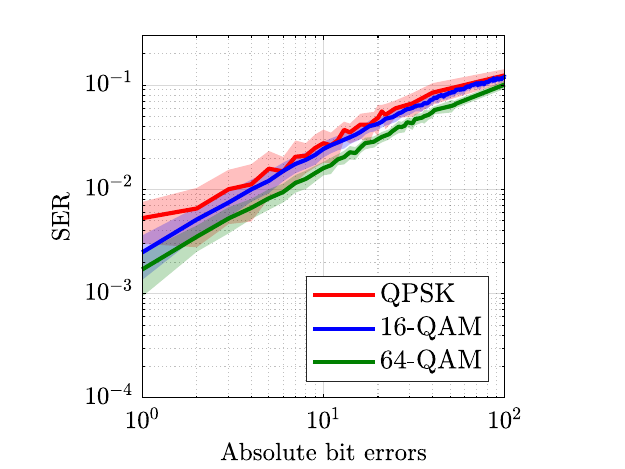}
		\caption{$M_{\mathrm{pl}}=512$, $r=2/3$}
		\label{fig:falModsymbModordnung}
	\end{subfigure}
	
	\caption[]{Interaction between information bit errors after decoding and SER in the reconstructed OFDM frame for different parameterizations. The solid lines represent the mean values and the standard deviations are drawn in transparent.}
	\label{fig:BERfalscheModsymbole}
\end{figure}

As described in Sec.~\ref{sec:RadarProcessing}, the full-frame radar image requires error-free  communication. Although channel coding significantly reduces the probability of bit errors, they still cannot be ruled out completely. Therefore, this section examines how erroneous data transmission affects the full-frame radar image.

Faulty communication automatically leads to incorrect modulation symbols when reconstructing the transmitted OFDM frame $\mathbf{\hat{X}}^{\mathrm{pl}}$. As a result, the elementwise division is not performed with the actual transmitted modulation symbols. In this context, Fig.~\ref{fig:BERfalscheModsymbole} illustrates the interaction between information bit errors after decoding and symbol error rate (SER) in the reconstructed OFDM frame for different parametrizations. Thus, the SER refers to the incorrectly reconstructed modulation symbols in $\mathbf{\hat{X}}^{\mathrm{pl}}$ compared to the originally transmitted OFDM frame. Since this is a radar investigation, only the current OFDM frame is considered. Therefore, the number of payload bits, and thus the achievable BER, varies depending on the code rate, frame size, and modulation order. Consequently, the SER is shown in terms of absolute bit errors instead of BER. For the case without channel coding, Fig.~\ref{fig:falModsymbuncodiert} shows a linear behavior between bit errors and SER. For small SER and Gray coding, the SER can be expressed according to \cite{chap4-cioffi} as
\begin{align}
\mathit{\mathrm{SER}} \approx \mathit{\mathrm{BER}} \cdot \log_{\mathrm{2}}(\mathcal{M}), \label{SER}
\end{align}
where BER represents the bit error rate and $\mathcal{M}$ is the modulation order. For QPSK modulation, $\mathcal{M}=4$. Essentially, this means that one bit error generates one incorrectly reconstructed modulation symbol, with the relationship between BER and SER given in \eqref{SER}. However, this changes when channel coding is used as can be seen from Figs.~\ref{fig:falModsymbCoderate}, \ref{fig:falModsymbFrame}, and \ref{fig:falModsymbModordnung}. In this case, the SER in the reconstructed OFDM frame can increase significantly. Even one incorrectly decoded bit has an enormous influence on the re-encoding process while reconstructing the transmitted OFDM frame. This results in a severely different bit sequence after re-encoding. Thus, one bit error before channel re-encoding results in many bit errors after channel re-encoding. This is because incorrect parity check bits are generated in addition to the original erroneous bit. Finally, this leads to the large number of incorrectly reconstructed modulation symbols.

As can be seen in Fig.~\ref{fig:falModsymbCoderate}, the influence of bit errors also increases as the code rate decreases. This follows from the fact that lower code rates lead to more parity check bits for each information bit, resulting in more erroneous bits after re-encoding. Besides, the plotted standard deviation shows that the exact SER for a given number of bit errors varies. Despite the perceived disadvantage of coded bits, it is still advantageous to use channel coding as it significantly reduces the probability of bit errors. It should also be emphasized that the different values of absolute bit errors in Fig.~\ref{fig:BERfalscheModsymbole} result from varying the SNR during the simulations. This is done to obtain a realistic bit error distribution and to include the influence of the SNR on the quality of synchronization, channel estimation, and equalization.

Looking at Figs.~\ref{fig:falModsymbFrame} and \ref{fig:falModsymbModordnung}, and assuming the same number of absolute bit errors, it is noticeable that the SER decreases when using a larger OFDM frame and higher modulation order. With larger OFDM frames, this can be verified mathematically, as the impact of the same number of incorrect modulation symbols is more pronounced in a smaller OFDM frame compared to a larger one, given that the percentage of symbol errors is higher in the former. Furthermore, with higher modulation orders, more wrong bits can be combined into one modulation symbol, resulting in fewer incorrectly reconstructed modulation symbols for the same number of bit errors. However, the OFDM frame and the modulation order cannot be increased arbitrarily. The number of transmitted OFDM symbols is limited by the measurement time and the bandwidth is limited by the sampling frequency or legal requirements. Furthermore, with higher modulation orders, higher mean constellation energy, and therefore SNR, is required to achieve the same BER. In addition, Tab.~\ref{table:BitfehlersalscheModsymbWerte} lists the mean SER for the different parametrizations at one and ten bit errors.

\begin{figure}[t!]
	\centering
	\begin{subfigure}[b]{0.206\textwidth}
		\centering
		\includegraphics[trim = 0cm 0cm 0cm 0cm,clip,width=\textwidth]{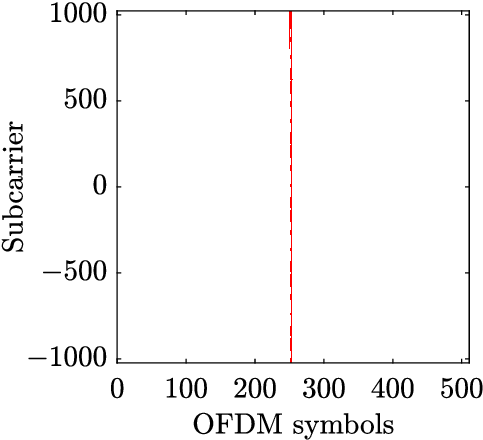}
		\caption{Reconstructed OFDM frame\\\ }
		\label{fig:rekFrame}
	\end{subfigure}
	\hfill
	\begin{subfigure}[b]{0.275\textwidth}
		\centering
		\includegraphics[trim = 0cm 0cm -3.3cm 0cm,clip,width=\textwidth]{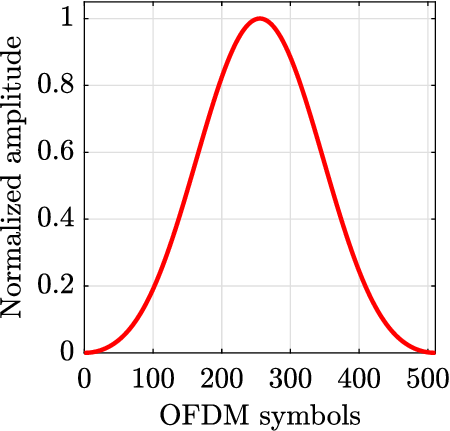}
		\caption{Blackman window in OFDM symbol\\direction}
		\label{fig:BlackmanFenster}
	\end{subfigure}
	\par\bigskip
	\begin{subfigure}[b]{0.208\textwidth}
		\includegraphics[trim = -0.34cm 0.08cm 2.75cm 0.3cm,clip,width=\textwidth]{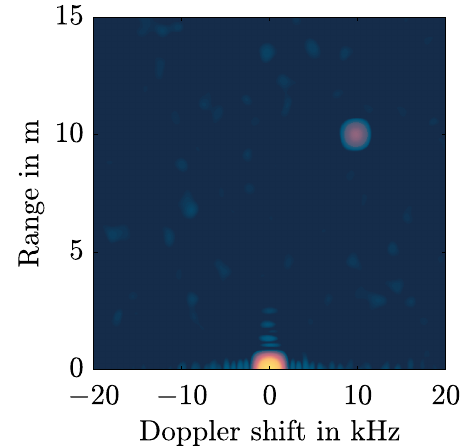}
		\caption{Radar image for block at the frame edge}
		\label{fig:RadarbildRand}
	\end{subfigure}
	\hfill
	\begin{subfigure}[b]{0.255\textwidth}
		\includegraphics[trim = 1cm 0.08cm 0.52cm 0.3cm,clip,width=0.9\textwidth]{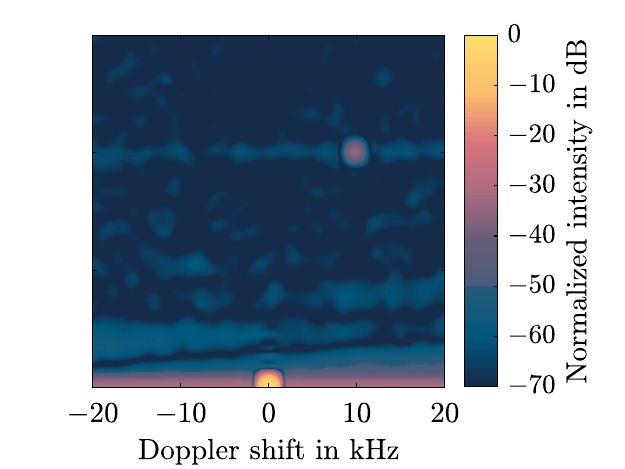}
		\caption{Radar image for block in the frame\\ center \textcolor{white}{}}
		\label{fig:RadarbildMitte}
	\end{subfigure}
	\caption{Influence of one bit error on the full-frame radar image for $M_{\mathrm{pl}}=512$, $r=2/3$, and a QPSK modulation. In (a) correctly and incorrectly reconstructed modulation symbols are shown in white and red, respectively.}
	\label{fig:EinflussBitfheler}
\end{figure}

Next, Fig.~\ref{fig:EinflussBitfheler} shows the influence of one bit error on the full-frame radar image for $M_{\mathrm{pl}}=512$, $r=2/3$ and QPSK modulation. As can be seen in Fig.~\ref{fig:rekFrame}, incorrectly reconstructed modulation symbols are arranged in a block structure in subcarrier direction. This is due to the facts that so many wrong modulation symbols are generated while reconstructing the transmitted OFDM frame and that the modulation symbols are inserted into the OFDM frame in column direction. However, the exact position of this block within the OFDM frame determines whether the full-frame radar image is affected or not. For blocks located at the frame edges, the interference caused by the incorrect modulation symbols is suppressed by the used Blackman window in Fig.~\ref{fig:BlackmanFenster}. Consequently, the resulting full-frame radar image in Fig.~\ref{fig:RadarbildRand} is not significantly disturbed. If, on the other hand, the block is located in the center of the OFDM frame, the interference is not suppressed. As can be seen in Fig.~\ref{fig:RadarbildMitte}, this results in the formation of stripes around the peaks in Doppler direction and the noise floor increases. The intensity of these stripes also depends on the strength of the peaks.

\begin{table}[t!]
	\renewcommand{\arraystretch}{1.5}
	\arrayrulecolor[HTML]{000000}
	\setlength{\arrayrulewidth}{.1mm}
	\setlength{\tabcolsep}{4pt}
	
	\centering
	\captionsetup{width=43pc,justification=centering,labelsep=newline}
	\caption{Mean SER in the reconstructed OFDM frame for different parametrizations\\ at one and ten bit errors}
	\resizebox{\columnwidth}{!}{
		\begin{tabular}{|c|c|c|}
			\hhline{|===|}
			Parametrization  & One bit error & Ten bit errors \\
			\hhline{|===|}
			$M_{\mathrm{pl}}=512$, uncoded, QPSK & $9.5\cdot 10^{-7}$ & $9.5\cdot 10^{-6}$ \\ 
			\cline{1-3}
			$M_{\mathrm{pl}}=512$, $r=1/3$, QPSK & $1.06\cdot 10^{-2}$ & $6.77\cdot 10^{-2}$ \\
			\cline{1-3}
			$M_{\mathrm{pl}}=512$, $r=2/3$, QPSK & $5.30\cdot 10^{-3}$ & $2.77\cdot 10^{-2}$ \\
			\cline{1-3}
			$M_{\mathrm{pl}}=512$, $r=5/6$, QPSK & $2.64\cdot 10^{-3}$ & $2.16\cdot 10^{-2}$ \\
			\cline{1-3}
			$M_{\mathrm{pl}}=1024$, $r=2/3$, QPSK & $2.50\cdot 10^{-3}$ & $1.54\cdot 10^{-2}$ \\
			\cline{1-3}
			$M_{\mathrm{pl}}=4096$, $r=2/3$, QPSK & $7.04\cdot 10^{-4}$ & $3.06\cdot 10^{-3}$ \\
			\cline{1-3}
			$M_{\mathrm{pl}}=512$, $r=2/3$, 16-QAM & $2.49\cdot 10^{-3}$ & $2.44\cdot 10^{-2}$ \\
			\cline{1-3}
			$M_{\mathrm{pl}}=512$, $r=2/3$, 64-QAM & $1.71\cdot 10^{-3}$ & $1.60\cdot 10^{-2}$ \\
			\hhline{|===|}
		\end{tabular}
	}
	\label{table:BitfehlersalscheModsymbWerte}
\end{table}

\begin{figure}[t!]
	\centering
	\begin{subfigure}[b]{0.206\textwidth}
		\centering
		\includegraphics[trim = 0cm 0cm 0cm 0cm,clip,width=\textwidth]{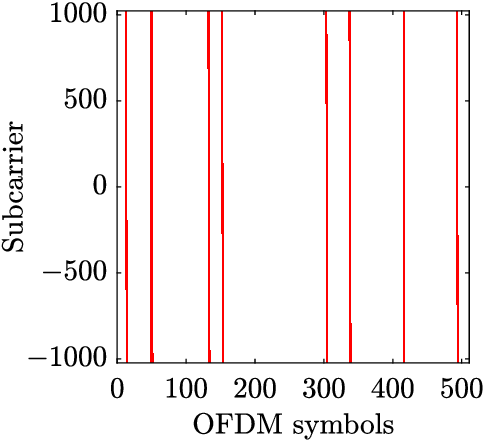}
		\caption{Irregular block arrangement\\\ }
		\label{fig:falscheModdaten_ungleichmaessig}
	\end{subfigure}
	\hfill
	\begin{subfigure}[b]{0.262\textwidth}
		\includegraphics[trim = 0cm 0.08cm 0.25cm 0.3cm,clip,width=\textwidth]{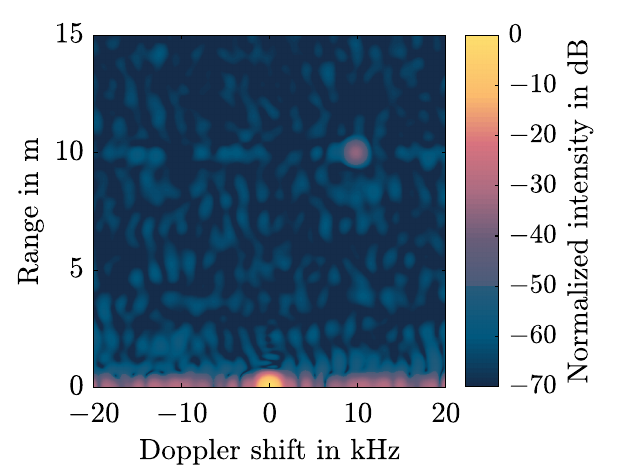}
		\caption{Full-frame radar image for irregular block arrangement}
		\label{fig:Radarbild_ungleichmaessig}
	\end{subfigure}
	\par\bigskip
	\begin{subfigure}[b]{0.206\textwidth}
		\centering
		\includegraphics[trim = 0cm 0cm 0cm 0cm,clip,width=\textwidth]{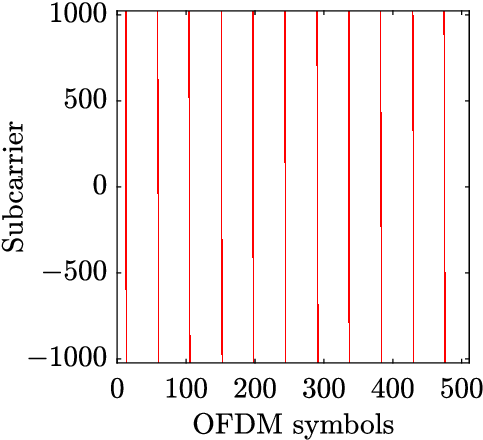}
		\caption{Regular block arrangement\\\ }
		\label{fig:falscheModdaten_gleichmaessig}
	\end{subfigure}
	\hfill
	\begin{subfigure}[b]{0.262\textwidth}
		\includegraphics[trim = 0cm 0.08cm 0.25cm 0.3cm,clip,width=\textwidth]{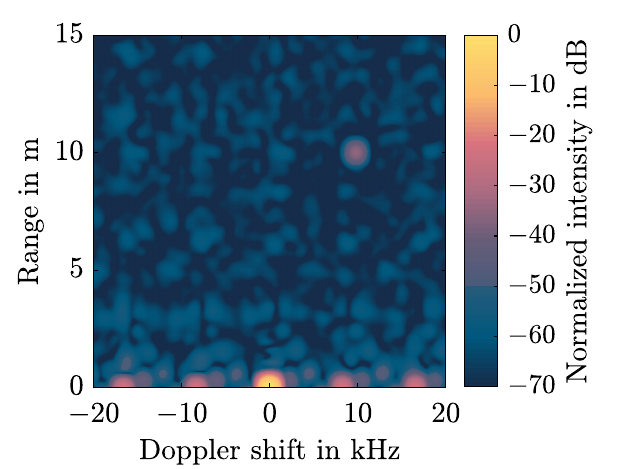}
		\caption{Full-frame radar image for regular block arrangement}
		\label{fig:Radarbild_gleichmaessig}
	\end{subfigure}
	\caption{Influence of several blocks with wrong modulation symbols on the full-frame radar image for $SER=0.05$. In (a) and (c) correctly and incorrectly reconstructed modulation symbols are shown in white and red, respectively.}
	\label{fig:mehrereBloecke}
\end{figure}

Next, Fig.~\ref{fig:mehrereBloecke} extends the investigations to the consideration of several blocks with wrong modulation symbols. Fig.~\ref{fig:falscheModdaten_ungleichmaessig} shows the case where the blocks are arranged irregularly. As illustrated in Fig.~\ref{fig:Radarbild_ungleichmaessig}, this again leads to stripes and a noise floor enhancement in the full-frame radar image. If, on the other hand, the blocks are regularly distributed in the OFDM frame, ghost targets occur. This is illustrated in Figs.~\ref{fig:falscheModdaten_gleichmaessig} and \ref{fig:Radarbild_gleichmaessig}. The ghost targets appear to the left and right of the actual peak and are arranged in constant distances in Doppler direction. The Doppler frequency distance between the ghost targets is related to the time interval after which blocks with correct modulation symbols are repeated. In Fig.~\ref{fig:falscheModdaten_gleichmaessig}, this time interval is \SI{118.53}{\micro\second}. This results in a Doppler frequency spacing of $1/\SI{118.53}{\micro\second}=\SI{8.4}{\kilo\hertz}$, which is consistent with Fig.~\ref{fig:Radarbild_gleichmaessig}.

From the above investigations it can be concluded that even one bit error can render the full-frame radar image unusable. Therefore, no bit errors are tolerable for creating the full-frame radar image. As a result, there are two options for handling bit errors. The first possibility is to create the full-frame radar image only when communication is error-free, which can be verified with channel coding. If, instead, bit errors occur, the full-frame radar image is discarded and targets are detected based on the pilot radar image.

\begin{figure}[t!]
	\centering
	\begin{subfigure}[b]{0.206\textwidth}
		\centering
		\includegraphics[trim = 0cm 0cm 0cm 0cm,clip,width=\textwidth]{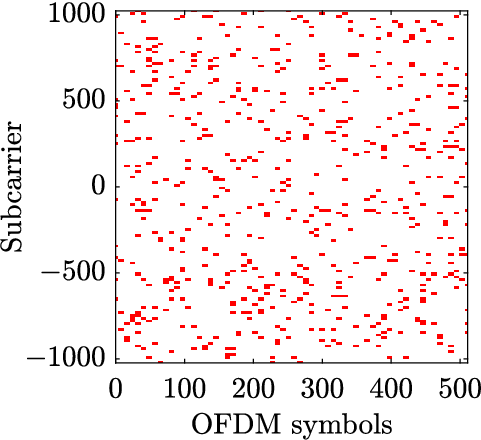}
		\caption{Random distribution for one bit error}
		\label{fig:rekFrame_random}
	\end{subfigure}
	\hfill
	\begin{subfigure}[b]{0.262\textwidth}
		\includegraphics[trim = 0cm 0.08cm 0.25cm 0.3cm,clip,width=\textwidth]{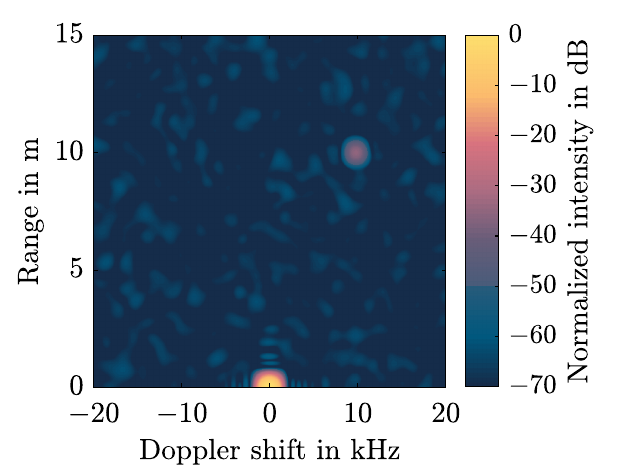}
		\caption{Full-frame radar image for one bit error\\\ }
		\label{fig:RadarbildRand_random}
	\end{subfigure}
	\par\bigskip
	\begin{subfigure}[]{0.5\textwidth}
		\includegraphics[trim = 0cm 0cm 0cm 0cm,clip,width=0.98\textwidth]{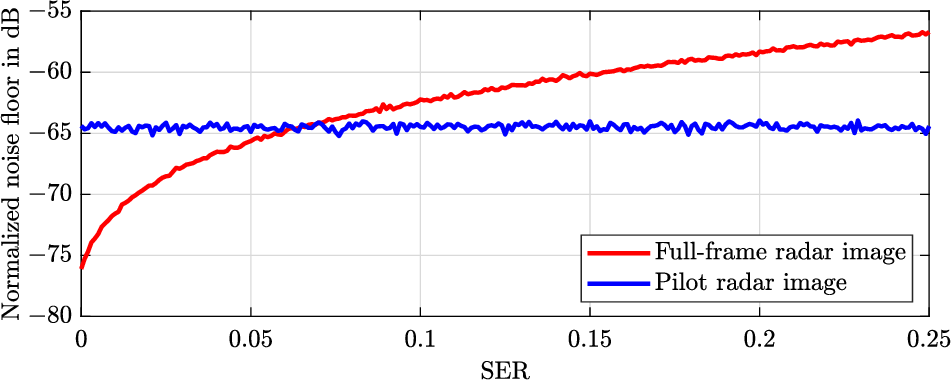}
		\caption{Noise floor depending on the SER per OFDM frame}
		\label{fig:RadarbildMitte_random}
	\end{subfigure}
	\caption[]{Random distribution of modulation symbols in the OFDM frame for $M_{\mathrm{pl}}=512$, $r=2/3$ and a QPSK modulation. In (a) correctly and incorrectly reconstructed modulation symbols are shown in white and red, respectively.}
	\label{fig:EinflussBitfheler_random}
\end{figure}

The second possibility is based on the idea of preventing the formation of stripes and ghost targets by resolving the block structure of the incorrectly reconstructed modulation symbols. As described before, the block structure is created by many consecutive bit errors after re-encoding as well as by inserting the modulation symbols in column direction into the OFDM frame. A countermeasure for that is to perform modulation symbol interleaving, where the modulation symbols are inserted randomly into the OFDM frame at the transmitter. This way, the incorrectly reconstructed modulation symbols also distribute themselves randomly within the OFDM frame. It is worth highlighting that symbol interleaving is already done in cellular systems, resulting in the fact that cellular frame numerology is robust against noise and interference burst and therefore also suitable for sensing operation. The symbol interleaving principle is illustrated in Fig.~\ref{fig:rekFrame_random} for one bit error. The resulting full-frame radar image in Fig.~\ref{fig:RadarbildRand_random} shows neither stripes nor ghost targets. Due to the fact that the interference is now evenly distributed within the OFDM frame, the effect of the wrong modulation symbols is merely an increase of the noise floor and an associated degradation of the SNR in the full-frame radar image. Based on this, Fig.~\ref{fig:RadarbildMitte_random} presents the noise floor of the full-frame radar image compared to the pilot radar image as a function of the SER. Since the pilot radar image is independent of communications, its noise floor remains constant over the SER. The full-frame radar image is superior to the pilot radar image up to an SER of about $0.07$. Tab.~\ref{table:zufaelligtolerierbareBitfehler} ties in with this value and shows the maximum number of tolerable bit errors with a mean SER of $0.07$. For better classification, the corresponding bit error rate is also listed. Consequently, this method can better tolerate a few bit errors that may happen after channel decoding, reducing their influence on the radar signal processing performed based on the reconstructed transmit frame.

\begin{table}[t!]
	\renewcommand{\arraystretch}{1.5}
	\arrayrulecolor[HTML]{000000}
	\setlength{\arrayrulewidth}{.1mm}
	\setlength{\tabcolsep}{4pt}
	
	\centering
	\captionsetup{width=43pc,justification=centering,labelsep=newline}
	\caption{Tolerable bit errors for a mean SER of $0.07$ and different parametrizations}
	\resizebox{\columnwidth}{!}{
		\begin{tabular}{|c|c|c|}
			\hhline{|===|}
			Parametrization  & Tolerable number of bit errors & Tolerable bit error rate \\
			\hhline{|===|}
			$M_{\mathrm{pl}}=512$, uncoded, QPSK & $73400$ & $37.7\cdot 10^{-3}$\\
			\cline{1-3}
			$M_{\mathrm{pl}}=512$, $r=1/3$, QPSK & $11$ & $1.7\cdot 10^{-5}$\\
			\cline{1-3}
			$M_{\mathrm{pl}}=512$, $r=2/3$, QPSK &  $32$ & $2.5\cdot 10^{-5}$\\
			\cline{1-3}
			$M_{\mathrm{pl}}=512$, $r=5/6$, QPSK & $79$ & $4.9\cdot 10^{-5}$\\
			\cline{1-3}
			$M_{\mathrm{pl}}=1024$, $r=2/3$, QPSK & $69$ & $2.7\cdot 10^{-5}$\\
			\cline{1-3}
			$M_{\mathrm{pl}}=4096$, $r=2/3$, QPSK &  $233$ & $2.2\cdot 10^{-5}$\\
			\cline{1-3}
			$M_{\mathrm{pl}}=512$, $r=2/3$, 16-QAM & $39$ & $2.7\cdot 10^{-5}$\\
			\cline{1-3}
			$M_{\mathrm{pl}}=512$, $r=2/3$, 64-QAM & $57$ & $1.5\cdot 10^{-5}$\\
			\hhline{|===|}
		\end{tabular}
	}
	\label{table:zufaelligtolerierbareBitfehler}
\end{table}

For the present case with $M_{\mathrm{pl}}=512$, $r=2/3$, QPSK modulation, and one bit error, a mean SER of $5.30\cdot 10^{-3}$ can be read from Tab.~\ref{table:BitfehlersalscheModsymbWerte}. As shown in Fig.~\ref{fig:RadarbildMitte_random}, this leads to an increase of the noise floor by \SI{3}{dB} compared to the case where $\mathrm{SER} = 0$. However, this effect is less severe than in case where modulation symbol scrambling is not performed, leading to stripes in the Doppler shift direction as seen in Fig.~\ref{fig:RadarbildMitte}. Thus, targets can still be detected without having strong degradation in the full-frame radar image.

\subsection{Impact of Synchronization Offsets}\label{Impact_SynchOffsets}
\subsubsection{Time Offset}
As a result of the spatial separation of transmitter and receiver, TO, FO, and SFO occur, which must be corrected at the receiver. However, their determination is only based on estimates, which potentially leads to residual offsets even after synchronization. Therefore, this section examines how synchronization offsets affect the ISAC system.

\begin{figure}[t!]
	\centering
	\begin{subfigure}[b]{0.23\textwidth}
		\centering
		\includegraphics[trim = 4.4cm 0.05cm 4cm 0.3cm,clip,width=0.965\textwidth]{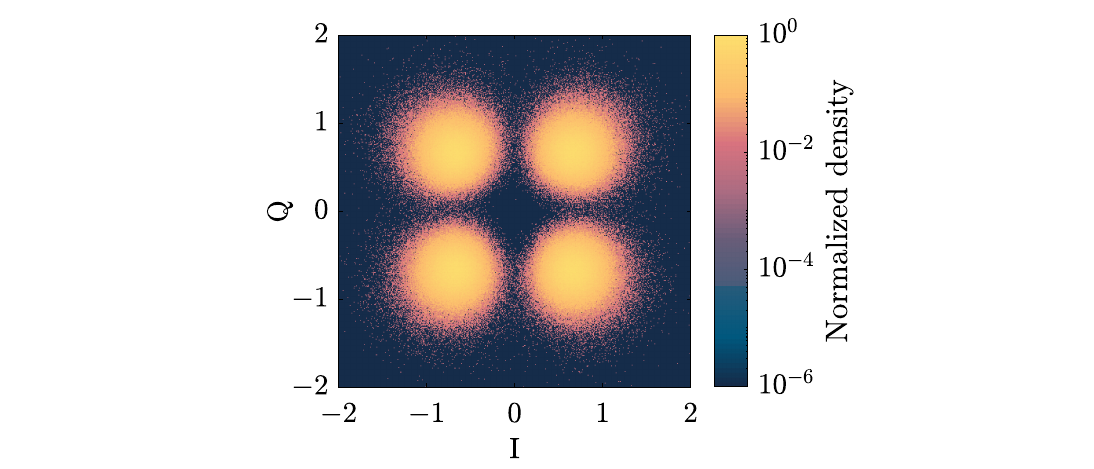}
		\caption{Noise-like broadening}
		\label{fig:KDTO}
	\end{subfigure}
	\hfill
	\begin{subfigure}[b]{0.252\textwidth}
		\includegraphics[trim = 0.3cm 0.08cm 0cm 0.3cm,clip,width=0.9\textwidth]{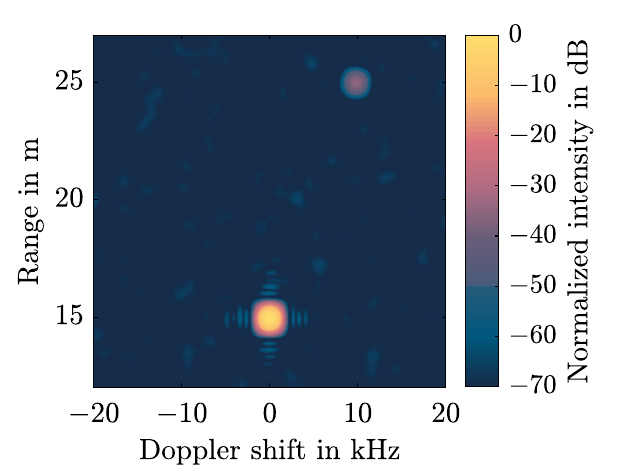}
		\caption{Shift in range direction}
		\label{fig:RadarbildTO}
	\end{subfigure}
	
	\caption{Effects of TO on the constellation diagram in (\subref{fig:KDTO}) and the radar image in~(\subref{fig:RadarbildTO}). A TO of $\tau_{\mathrm{off}} = 100 \text{ samples}$ was used for simulation and all \mbox{$M_\mathrm{pl}=512$} OFDM symbols were taken into account.}
	\label{fig:TOPrinzip}
\end{figure}

\begin{figure}[t!]
	\centering
	\begin{subfigure}[b]{0.199\textwidth}
		\includegraphics[width=\textwidth]{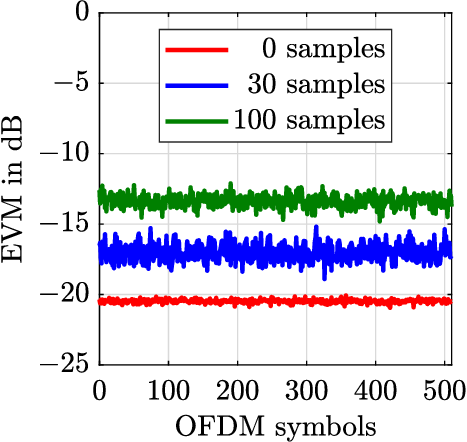}
		\caption{EVM over OFDM symbols}
		\label{fig:TO_Symbole}
	\end{subfigure}
	\hspace{0.5cm}
	\begin{subfigure}[b]{0.205\textwidth}
		\includegraphics[width=\textwidth]{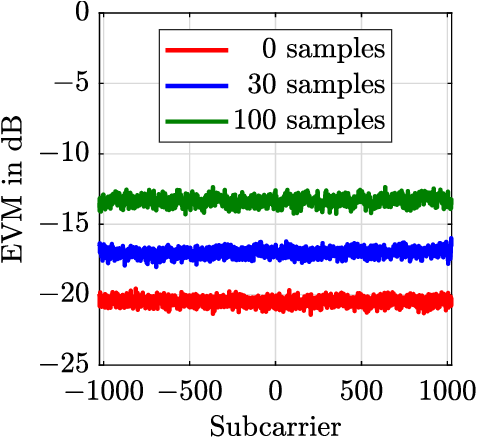}
		\caption{EVM over subcarrier}
		\label{fig:TO_Untertraeger}
	\end{subfigure}
	
	\caption{Investigation of the EVM in presence of TO. In (a) the EVM is averaged over the used subcarriers and in (b) over the transmitted OFDM symbols.}
	\label{fig:EVM_TO}
\end{figure}

In case of perfect time synchronization, the evaluation range of the DFT window is exactly aligned with the original OFDM symbol. A time shift of the start point of the DFT window within the CP leads to a frequency-dependent phase shift of the subcarriers, which, however, can be compensated to a certain extent by equalization. However, as soon as the starting point of the DFT window exceeds the CP, ISI occurs. As can be seen from Fig.~\ref{fig:KDTO}, this leads to a noise-like broadening of the constellation diagram and an associated deterioration of the EVM. This deterioration of the EVM is constant over the OFDM symbol and subcarrier indices, as Figs.~\ref{fig:TO_Symbole} and \ref{fig:TO_Untertraeger} illustrate.

Any TO, whether inside or outside the CP, changes the propagation time of the signals, causing the radar image to shift in range direction. This is shown in Fig.~\ref{fig:RadarbildTO}. In case ISI occurs, the SNR in the radar image additionally decreases.

\subsubsection{Frequency Offset}
With an FO, the subcarriers of the received signal shift in comparison to the ideal frequency points. Hence, all subcarriers experience the same frequency shift. This causes the OFDM symbol to be sampled at the wrong frequency points, leading to a loss of orthogonality. The result is a decrease in signal amplitude as well as ICI due to the mutual influence of the subcarriers.

\begin{figure}[t!]
	\centering
	\begin{subfigure}[b]{0.23\textwidth}
		\centering
		\includegraphics[trim = 4.4cm 0.08cm 4cm 0.3cm,clip,width=0.965\textwidth]{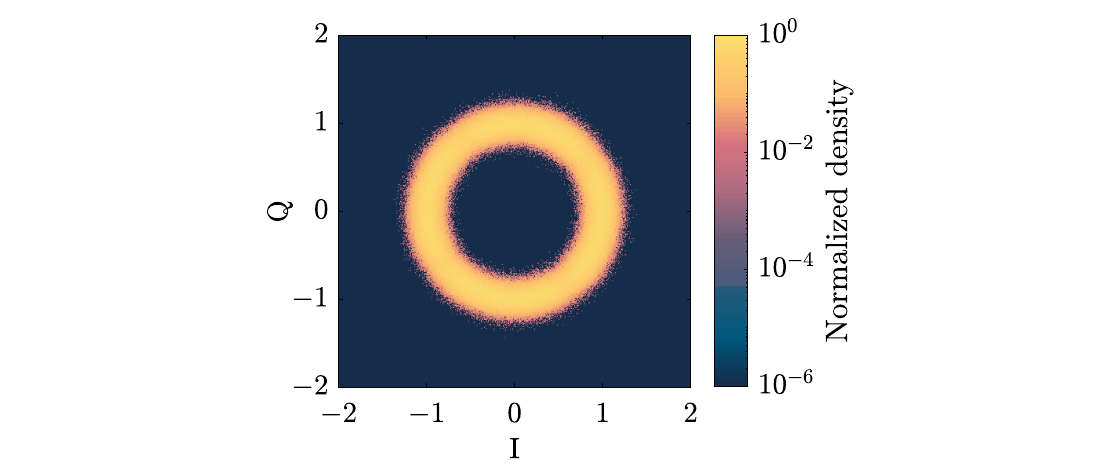}
		\caption{Arrangement on a circular path}
		\label{fig:KDFO}
	\end{subfigure}
	\hfill
	\begin{subfigure}[b]{0.25\textwidth}
		\includegraphics[trim = 0.3cm 0.05cm 0cm 0.3cm,clip,width=0.9\textwidth]{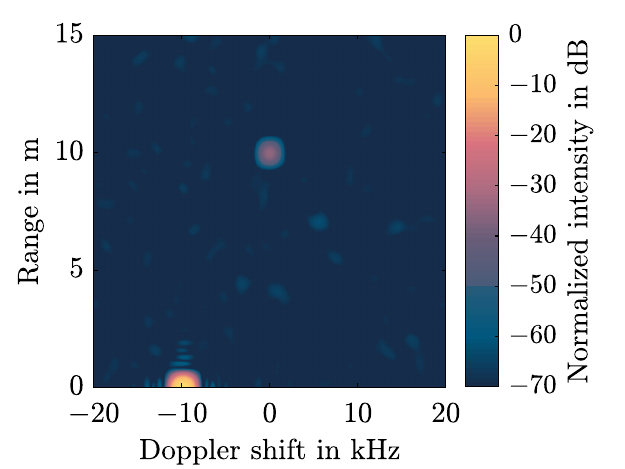}
		\caption{Shift in frequency direction}
		\label{fig:RadarbildFO}
	\end{subfigure}
	
	\caption{Effects of FO on the constellation diagram in (\subref{fig:KDFO}) and the radar image in~(\subref{fig:RadarbildFO}). An FO of $f_{\mathrm{off}} = \SI{-10}{\kilo\hertz}$ was used for simulation and all $M_\mathrm{pl}=512$ OFDM symbols were taken into account.}
	\label{fig:FOPrinzip}
\end{figure}

\begin{figure}[t!]
	\centering
	\begin{subfigure}[b]{0.199\textwidth}
		\includegraphics[width=\textwidth]{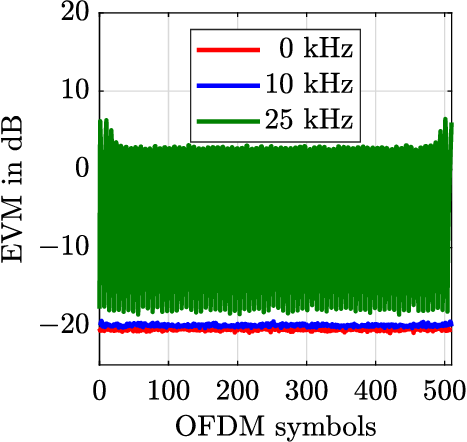}
		\caption{EVM over OFDM symbols}
		\label{fig:FO_Symbole}
	\end{subfigure}
	\hspace{0.5cm}
	\begin{subfigure}[b]{0.205\textwidth}
		\includegraphics[width=\textwidth]{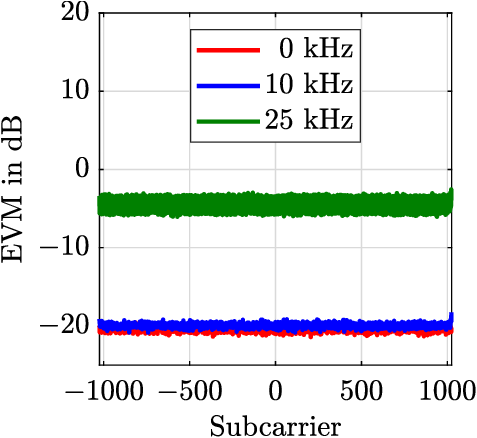}
		\caption{EVM over subcarrier}
		\label{fig:FO_Untertraeger}
	\end{subfigure}
	
	\caption{Investigation of the EVM in presence of FO. In (a) the EVM is averaged over the used subcarriers and in (b) over the transmitted OFDM symbols.}
	\label{fig:EVM_FO}
\end{figure}

The basic effect of an FO on the constellation diagram is shown in Fig.~\ref{fig:KDFO}. An FO results in a phase shift along different OFDM symbols, which causes the received modulation symbols to arrange themselves on a circular path in the constellation diagram. This phase shift can again be compensated to a certain extent by equalization. This is also illustrated by Figs.~\ref{fig:FO_Symbole} and \ref{fig:FO_Untertraeger}, where the EVM does not deteriorate for an FO of $\SI{10}{\kilo\hertz}$. However, if the FO becomes too large, the channel estimation will only be correct for the OFDM symbols containing pilots. In this case, \eqref{Mt} no longer applies and the OFDM frame does not contain enough pilots for proper channel estimation for the given FO. Consequently, the EVM oscillates over the OFDM symbol index with the period $M_{\mathrm{t}}$ and deteriorates overall. This is presented in Figs.~\ref{fig:FO_Symbole} and \ref{fig:FO_Untertraeger} as an example for an FO of \SI{25}{\kilo\hertz}.

As can be seen in Fig.~\ref{fig:RadarbildFO}, an FO causes a shift of the radar image in frequency direction. As long as the FO is small compared to the subcarrier spacing, the FO can be considered ICI-free according to Tab.~\ref{tab:radarParameters}. In this case, the radar image experiences only a slight SNR loss.

\subsubsection{Sampling Frequency Offset}
\begin{figure}[t!]
	\centering
	\begin{subfigure}[b]{0.23\textwidth}
		\includegraphics[width=\textwidth]{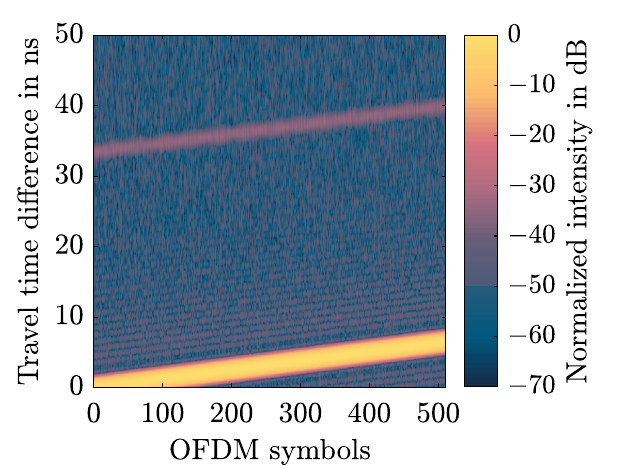}
		\caption{Travel time difference over OFDM symbols}
		\label{fig:SFO_DTP}
	\end{subfigure}
	\hfill
	\begin{subfigure}[b]{0.25\textwidth}
		\centering
		\includegraphics[trim = 0cm 0.05cm -0.9cm 0.3cm,clip,width=\textwidth]{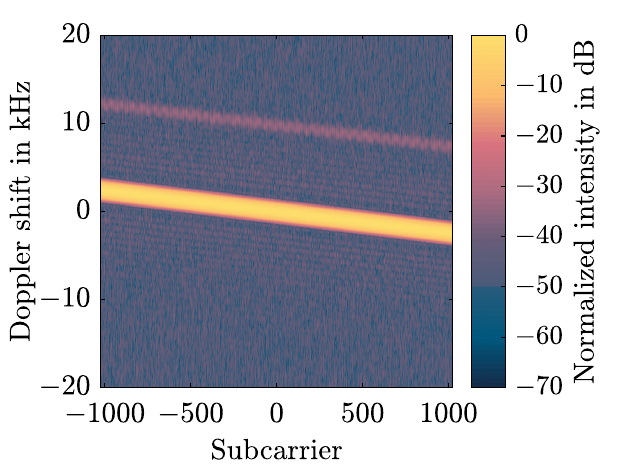}
		\caption{Doppler shift over subcarrier\\\ }
		\label{fig:SFO_DDP}
	\end{subfigure}
	\par \bigskip
	\begin{subfigure}[b]{0.23\textwidth}
		\centering
		\includegraphics[trim = 4.4cm 0.05cm 4cm 0.3cm,clip,width=0.965\textwidth]{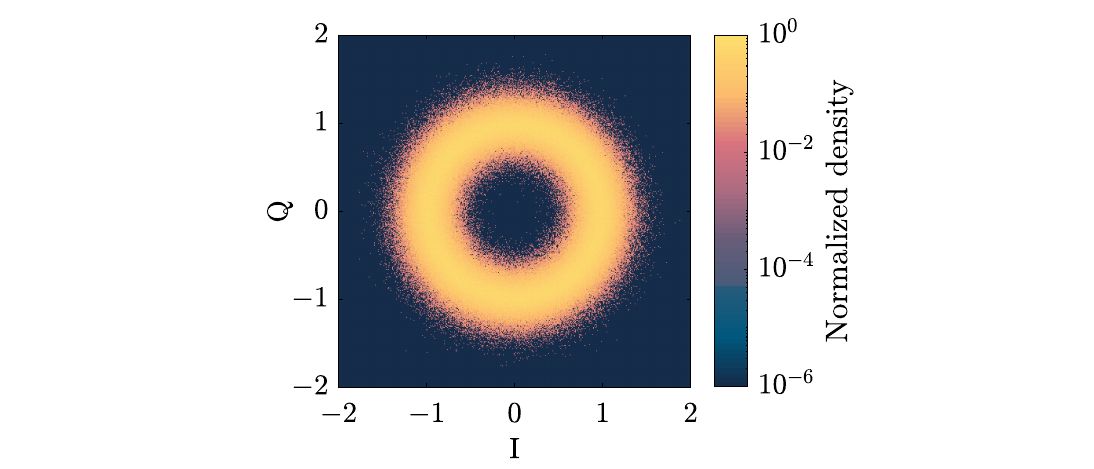}
		\caption{Noise-like broadening and\\ arrangement on a circular path}
		\label{fig:KDSFO}
	\end{subfigure}
	\begin{subfigure}[b]{0.245\textwidth}
		\includegraphics[trim = 0cm 0.05cm 0.2cm 0cm,clip,width=0.925\textwidth]{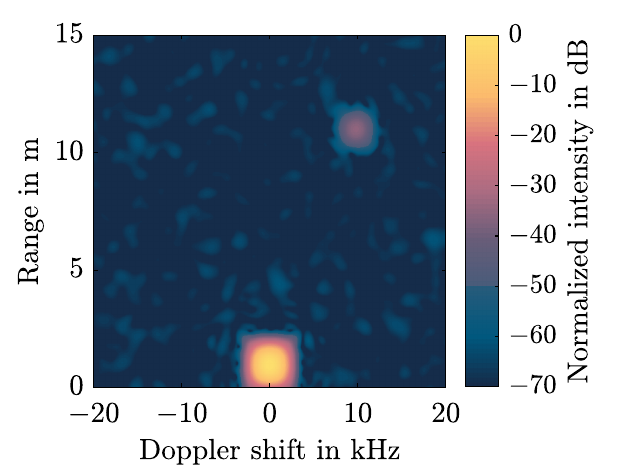}
		\caption{Range and Doppler migration and\\ shift in range direction}
		\label{fig:RadarbildSFO}
	\end{subfigure}
	
	\caption{Effects of SFO on the travel time difference in~(\subref{fig:SFO_DTP}), Doppler shift in~(\subref{fig:SFO_DDP}), constellation diagram in~(\subref{fig:KDSFO}) and radar image in~(\subref{fig:RadarbildSFO}). Different SFO values were used to better highlight the SFO effects. In~(\subref{fig:SFO_DTP}), (\subref{fig:SFO_DDP}) and (\subref{fig:RadarbildSFO}) $\delta_{\mathrm{off}} = \SI{5}{\text{ppm}}$ and in~(\subref{fig:KDSFO}) $\delta_{\mathrm{off}} = \SI{150}{\text{ppm}}$.}
	\label{fig:SFOPrinzip}
\end{figure}

\begin{figure}[t!]
	\centering
	\begin{subfigure}[b]{0.199\textwidth}
		\includegraphics[width=\textwidth]{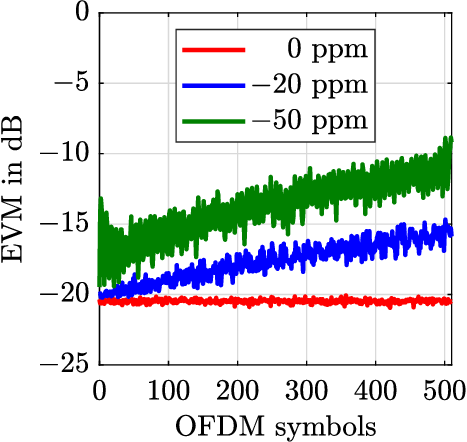}
		\caption{EVM over OFDM symbols}
		\label{fig:SFO_Symbole}
	\end{subfigure}
	\hspace{0.5cm}
	\begin{subfigure}[b]{0.205\textwidth}
		\includegraphics[width=\textwidth]{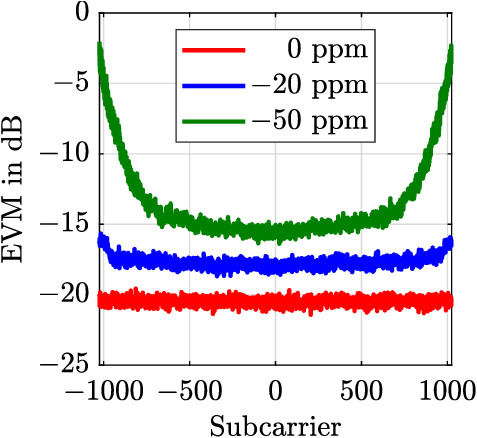}
		\caption{EVM over subcarrier}
		\label{fig:SFO_Untertraeger}
	\end{subfigure}
	
	\caption{Investigation of the EVM in presence of SFO. In (a) the EVM is averaged over the used subcarriers and in (b) over the transmitted OFDM symbols.}
	\label{fig:EVM_SFO}
\end{figure}

As illustrated in Figs.~\ref{fig:SFO_DTP} and \ref{fig:SFO_DDP}, an SFO causes a change in the travel time difference and Doppler shift as a function of the OFDM symbol and subcarrier index. This results in a time-dependent time error as well as a frequency-dependent frequency error. Thus, the SFO combines the effects of the TO and FO so that ISI, ICI, and a decrease in signal amplitude along the subcarriers occur simultaneously \cite{Smaini}. In addition, these effects increase with higher OFDM symbol and subcarrier indices, since time and frequency errors increase accordingly. The combination of TO and FO can also be seen in the constellation diagram in Fig.~\ref{fig:KDSFO}. The frequency error causes a phase shift in the constellation diagram, so that the received modulation symbols are again arranged on a circular path. However, due to the additional ISI, the circular path widens inwards and outwards. A more detailed investigation of the EVM is presented in Figs.~\ref{fig:SFO_Symbole} and \ref{fig:SFO_Untertraeger}. Due to its additive behavior, the SFO deteriorates the EVM with increasing OFDM symbol and subcarrier indices. Thus, bit errors tend to occur in the outer regions of the OFDM frame.

The effects of an SFO on the radar image are shown in Fig.~\ref{fig:RadarbildSFO}. The changing travel time difference and Doppler shift lead to range and Doppler shift migration, which results in a deterioration of both range and Doppler shift resolutions. Furthermore, the radar image experiences a slight shift in range direction. This is also due to the changing travel time difference, as it causes the signal power to center at an incorrect range. Accordingly, the signal power centers at the range corresponding to the mean travel time difference of the respective propagation path. With an SFO, the SNR decreases even at small offset values. This is related to ISI, ICI, and the decrease in signal amplitude along the subcarriers as well as range and Doppler shift migration, which spread the target reflection power over a larger area in the radar image.

\subsection{Synchronization Results}
\begin{figure}[t!]
	\centering
	\begin{subfigure}{0.21\textwidth}
		\centering
		\includegraphics[width=\textwidth]{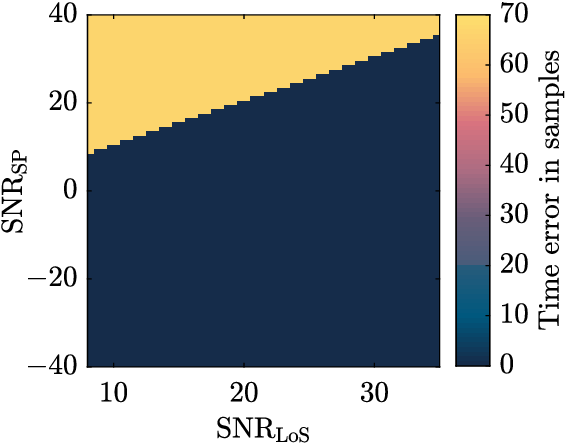}
		\caption{Absolute value of the mean time error}
		\label{fig:mittlererTO}
	\end{subfigure}
	\par \bigskip \smallskip
	\begin{subfigure}[b]{0.21\textwidth}
		\includegraphics[width=\textwidth]{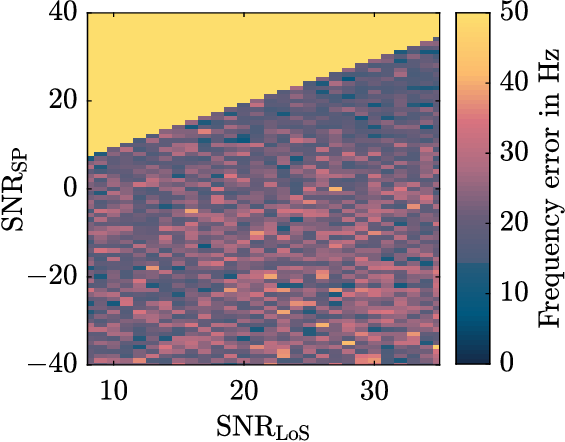}
		\caption{Absolute value of the mean frequency error}
		\label{fig:mittlererFO}
	\end{subfigure}
	\hspace{0.3cm}
	\begin{subfigure}[b]{0.222\textwidth}
		\centering
		\includegraphics[width=\textwidth]{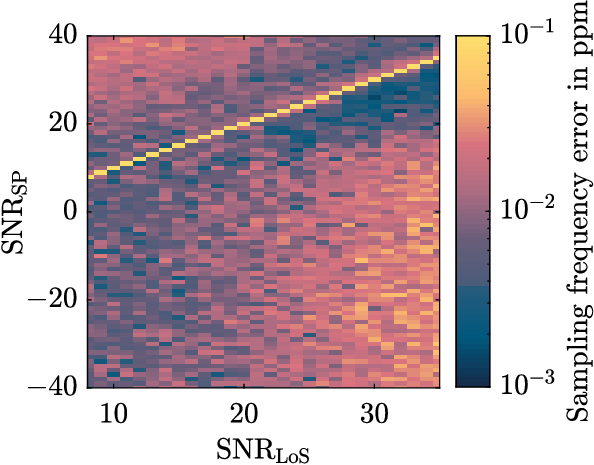}
		\caption{Absolute value of the mean sampling frequency error}
		\label{fig:mittlererSFO}
	\end{subfigure}
	
	\caption{Synchronization errors for different values of $\mathrm{SNR_{\mathrm{LoS}}}$ and $\mathrm{SNR_{\mathrm{SP}}}$. The synchronization errors refer to the LoS path.}
	\label{fig:SynchfehlerbzglSBR}
\end{figure}

Sec.~\ref{Impact_SynchOffsets} has demonstrated the effects of synchronization offsets on the communication and the radar application. In this context, this section shows simulations of how accurate the synchronization is and which remaining errors must be expected. For the following investigations, the initial values for TO, FO, and SFO were set to $67$ samples, \SI{100}{\kilo\hertz} and \SI{100}{\text{ppm}}, respectively.

Fig.~\ref{fig:SynchfehlerbzglSBR} represents the remaining time, frequency, and sampling frequency errors of the LoS path after synchronization. The ISAC system is examined for different values of $\mathrm{SNR_{\mathrm{LoS}}}$ and $\mathrm{SNR_{\mathrm{SP}}}$ and each matrix point is averaged over ten runs. $\mathrm{SNR_{\mathrm{LoS}}}$ and $\mathrm{SNR_{\mathrm{SP}}}$ describe the SNR of the LoS and secondary path at the receiver input. The intended area of application of the ISAC system is in the region of \mbox{$\mathrm{SNR_{\mathrm{LoS}}} > \mathrm{SNR_{\mathrm{SP}}}$}. As can be seen, TO, FO, and SFO can be accurately compensated in this region. Due to the cross-correlation, perfect time synchronization can be achieved at sample level. Furthermore, the remaining frequency and sampling frequency errors are in a range smaller than \SI{50}{\hertz} and \SI{0.1}{\text{ppm}}. These high accuracies come, especially at low $\mathrm{SNR_{\mathrm{LoS}}}$, from the residual FO and residual SFO corrections. Consequently, the errors remaining after synchronization are mainly dependent on the sampling frequency and the achievable resolution in the DTP and DDP.

\section{Proof-of-Concept Measurement Results}\label{sec:DemonstrationsmmWaveFrequencies}
\begin{figure}[t!]
	\centering	
	\begin{tikzpicture}[]
	\node[inner sep=0pt] at (0,0) {\includegraphics[width=0.485\textwidth]{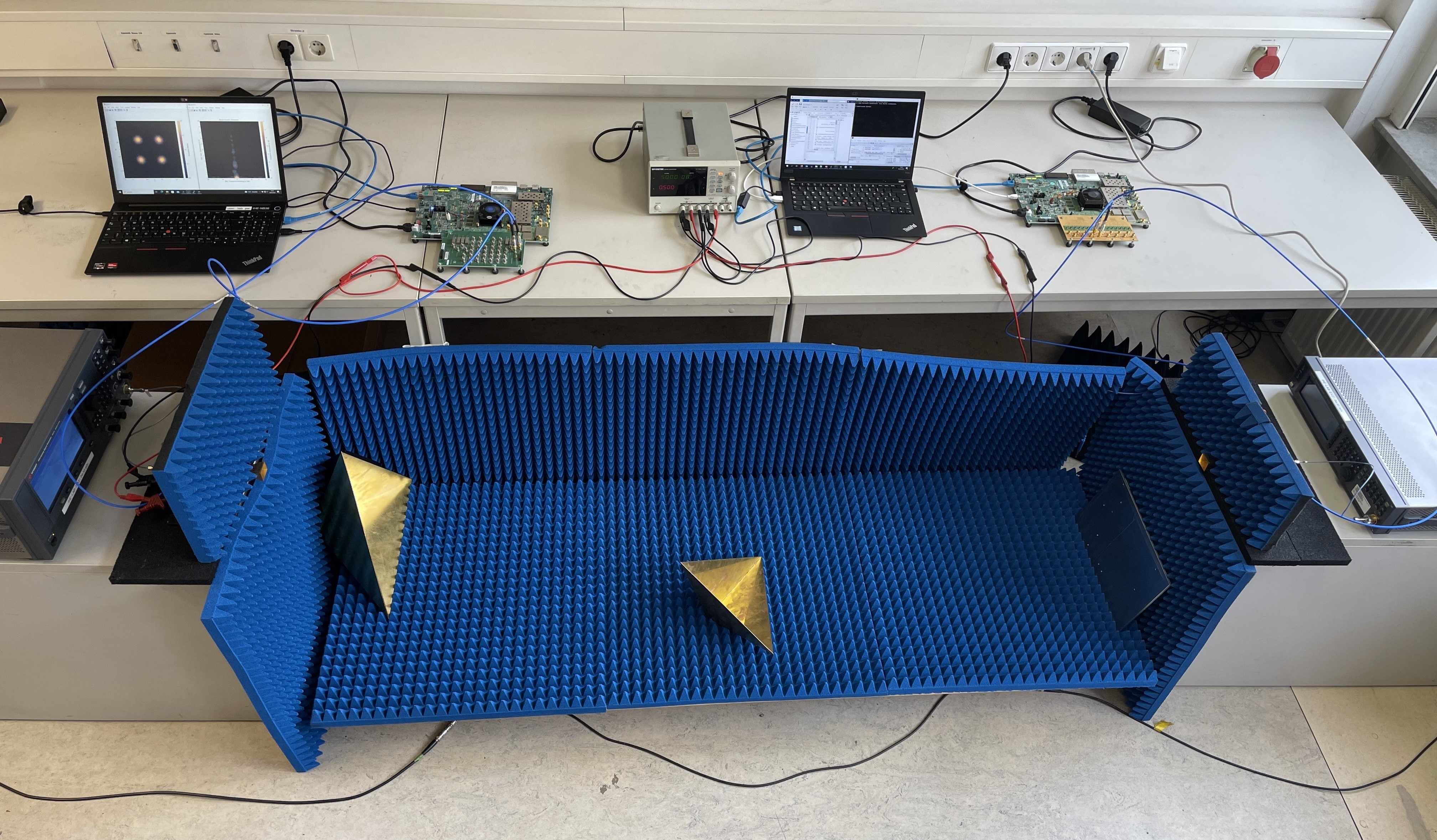}};
	\node [rectangle,draw,white,text = black,fill=white] at (-0.9,2.28) {\footnotesize Power supply};
	\node [rectangle,draw,white,text = black,fill=white] at (1.5,2.28) {\footnotesize Signal generation};
	\node [rectangle,draw,white,text = black,fill=white] at (-3.2,2.28) {\footnotesize Signal evaluation};
	
	\node [rectangle,draw,white,text = black,fill=white] at (3.6,-1.3) {\footnotesize Transmitter};
	\node [rectangle,draw,white,text = black,fill=white] at (-3.75,-1.3) {\footnotesize Receiver};
	
	\node [rectangle,draw,white,text = black,fill=white] at (-2.2,-1.9) {\footnotesize Target 1};
	\node [rectangle,draw,white,text = black,fill=white] at (0.2,-1.9) {\footnotesize Target 2};
	\node [rectangle,draw,white,text = black,fill=white] at (2.5,-1.9) {\footnotesize Reflector};
	
	\node [rectangle,draw,white,text = black,fill=white] at (2.3,1.75) {\footnotesize RFSoC};
	\node [rectangle,draw,white,text = black,fill=white] at (-1.5,1.75) {\footnotesize RFSoC};
	\end{tikzpicture}
	\caption{Measurement setup of the bistatic OFDM-based ISAC system at \SI{79}{GHz}}
	\label{figure:Messaufbau}
\end{figure}
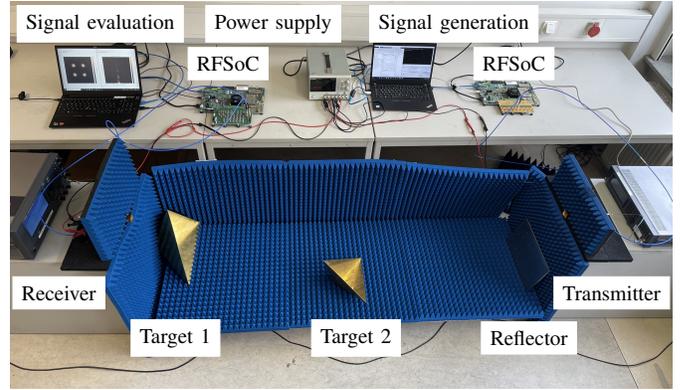

In this section, the bistatic OFDM-based ISAC system is verified via proof-of-concept measurement results at \SI{79}{\giga\hertz}. Although this frequency is not covered by cellular communication standards, it lies in the $\SI{77}{}-\SI{81}{\giga\hertz}$ short range radar (SRR) band defined in the standard ETSI EN 302 264 by the European Telecommunications Standards Institute (ETSI), which can potentially support ISAC operation for automotive applications in the future. The measurement system setup is shown in Fig.~\ref{figure:Messaufbau}. On the transmitting side, the complex OFDM baseband signal is generated on a host laptop. The parametrization of the OFDM frame is carried out according to Tab.~\ref{tab:resultsParameters}. However, the number of transmitted symbols is increased to $M_{\mathrm{pl}}=4096$ to demonstrate feasibility with larger OFDM frames. This changes the following communication and radar performance parameters: $\mathcal{R}_\mathrm{comm}=\SI{0.99}{Gbit/s}$, $G_\mathrm{p}=\SI{57.78}{dB}$ (pilots only), $G_\mathrm{p}=\SI{69.24}{dB}$ (full-frame), and $\Delta f_\mathrm{D}=\SI{95.37}{\hertz}$. Next, the OFDM signal is forwarded to a Zynq UltraScale+ RFSoC ZCU111 platform from Xilinx, where it is digitally upconverted to an intermediate frequency of \SI{1}{\giga\hertz}. Subsequently, the I and Q components undergo D/A conversion and are mixed up to \SI{79}{\giga\hertz} with an IQ mixer. The required LO signal is provided by a signal generator. After amplification, the signal is radiated with an antenna. On the receiving side, the signal is amplified and mixed down to \SI{1}{\giga\hertz} with an IQ mixer. The LO signal for down conversion is generated by a second signal generator. Next, the received signal is passed on to a second RFSoC platform, where it is digitized and converted to complex baseband to ultimately be evaluated.

Since the two RFSoC platforms use different reference clocks and are not synchronized with each other, STO, CFO, and SFO occur. The CFO results from the digital up-conversion stages. An additional CFO due to analog mixing is eliminated by synchronization between the used signal generators, which would be similar to using GPS-disciplined oscillators at both transmitter and receiver. For the radar application, two targets are placed in the measurement setup. As metioned in Sec.~\ref{sec:RadarProcessing}, the range determination is based on measuring the travel time difference between a secondary path and the LoS path. In order for the targets to be clearly visible in the radar image, the secondary paths must be recognizably longer than the LoS path. Therefore, the secondary paths are routed to the receiver via a reflector. Consequently, there are mainly three propagation paths:
\begin{itemize}
	\item LoS path: Tx $\rightarrow$ Rx
	\item Secondary path 1: Tx $\rightarrow$ Target 1 $\rightarrow$ Reflector $\rightarrow$ Rx
	\item Secondary path 2: Tx $\rightarrow$ Target 2 $\rightarrow$ Reflector $\rightarrow$ Rx
\end{itemize}
In order to also obtain a target with Doppler shift, the second target is additionally moved manually.

\begin{figure}[t!]
	\centering
	\begin{subfigure}[b]{0.24\textwidth}
		\includegraphics[trim = 4.45cm 0.08cm 3.5cm 0.3cm,clip,width=0.978\textwidth]{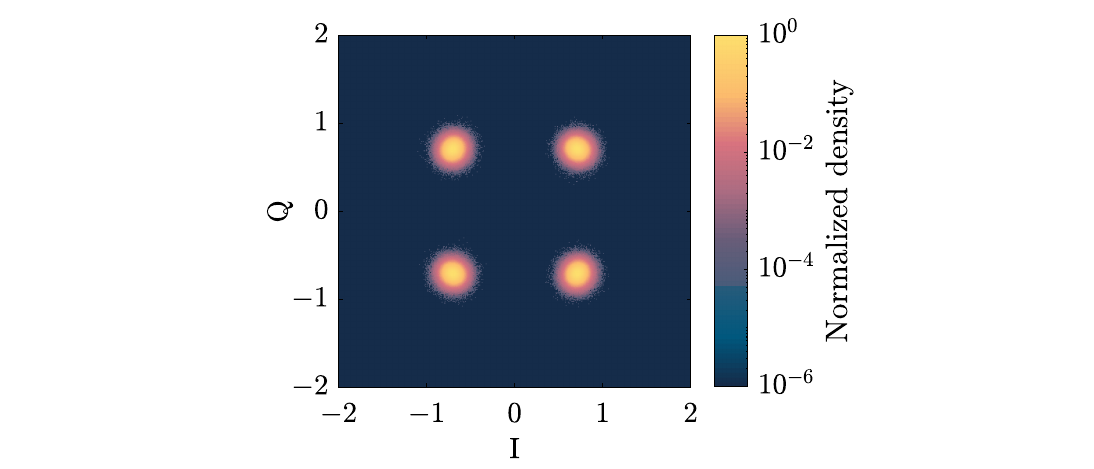}
		\caption{Constellation diagram}
		\label{fig:KD79GHz}
	\end{subfigure}
	\hfill
	\begin{subfigure}[b]{0.212\textwidth}
		\includegraphics[width=\textwidth]{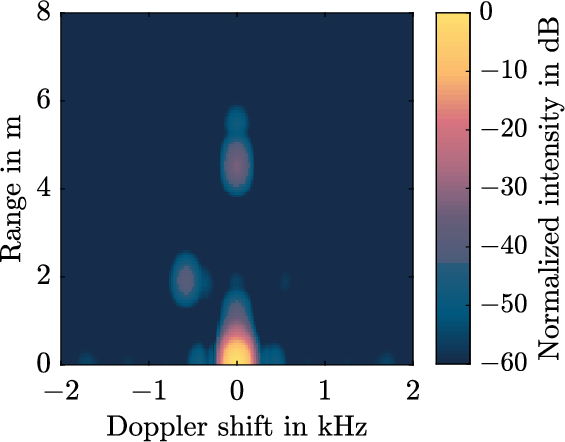}
		\caption{Full-frame radar image}
		\label{fig:FFRadarbild79GHz}
	\end{subfigure}
	
	\caption{Measurement result of the bistatic OFDM-based ISAC systems at \SI{79}{GHz}}
	\label{fig:ErgebnisMessung79GHz}
\end{figure}

The measurement results are shown in Fig.~\ref{fig:ErgebnisMessung79GHz}. As can be seen from the constellation diagram in Fig.~\ref{fig:KD79GHz}, all modulation symbols can be received correctly. Consequently, it is possible to create the full-frame radar image in Fig.~\ref{fig:FFRadarbild79GHz} without interference. Essentially, three peaks occur through the propagation paths named above. The peak at the lowest range represents the LoS path. The broadening in range direction is caused by additional reflections in the surroundings. These have a similar travel time to the LoS path and overlap with its peak. Furthermore, the peaks for the static target 1 and the moving target 2 are clearly visible. Since the measured LoS peak is at a relative bistatic range and Doppler shift of around \SI{0}{\meter} and \SI{0}{\kilo\hertz} and no range Doppler migration is formed, it can be conluded that the synchronization works accurately. The estimates for FO and SFO are \SI{-102,56}{\kilo\hertz} and \SI{102,56}{\text{ppm}}, respectively. Given the signal propagation times and the different switch-on times of transmitter and receiver, the TO takes random values for each measurement.

\section{Conclusions}\label{sec:Conclusion}
This article introduced a bistatic OFDM-based ISAC system with OTA synchronization that simultaneously allows transmitting binary data and detecting targets in the environment. In this context, estimation of TO, FO and SFO and their compensation were discussed, and bistatic radar sensing approaches based on the sole use of known pilot subcarriers at the receiver side or on the full transmit OFDM frame estimated via communication processing were analyzed. To avoid bit errors after decoding and therefore prevent degradation of the performance of both communication and sensing using the full frame for radar processing, FEC was assumed to be used.

The achieved results showed that bit errors after decoding causes visible degradation of the obtained bistatic radar images, eventually leading to ghost targets. To circumvent that, a random distribution of modulation symbols within the transmit OFDM frame was proposed to evenly spread the effect of data decoding failures on the radar image, solely leading to a slight noise level increase and therefore enhancing the robustness of the sensing capability in the bistatic OFDM-based ISAC system. Regarding synchronization mismatches, it was observed that the communication performance does not immediately deteriorate in presence of a TO and FO, but is significantly affected by SFO in long OFDM frames. As for sensing, TO and FO may bias range and Doppler shift estimates, respectively, and lead to SNR degradation in the radar image, while SFO combines both effects and also causes range and Doppler migration. This results in significant degradation of the radar sensing performance if uncorrected. Due to this reason, an SFO estimation and correction scheme consisting of a first estimation via preamble OFDM symbols, a residual estimation based on pilot subcarriers, and correction via resampling, was proposed.


\bibliographystyle{IEEEtran}
\bibliography{./main}

\end{document}